\pdfoutput=1

\documentclass[fleqn,usenatbib,useAMS]{mnras}

\usepackage{graphicx}	
\usepackage{amsmath}	
\usepackage{multicol}        
\usepackage{bm}		
\usepackage{pdflscape}	
\usepackage{latexsym}
\usepackage[authoryear]{natbib}
\usepackage{bm,nicefrac}
\usepackage{booktabs}

\usepackage[T1]{fontenc}
\usepackage{ae,aecompl}

\usepackage{newtxtext,newtxmath}

\def\deg{\ifmmode^\circ\else$^\circ$\fi}

\def\Q{\ifmmode\mathcal{Q}\else$\mathcal{Q}$\fi}
\def\Mach{\ifmmode\mathcal{M}\else$\mathcal{M}$\fi}

\title[The Hub-Filament System in G286.21+0.17]
{Discovery of a Compact Hub-Filament System in G286.21+0.17 with JWST and ALMA: Insights into Protocluster Formation and Competitive Accretion}
\author[L.~K. Dewangan et al.]{L.~K. Dewangan$^{1}$, 
N. K. Bhadari$^{2}$,
Ram~K. Yadav$^{3}$,
A.~K. Maity$^{1,4}$,
O.~R. Jadhav$^{1,5}$,
\newauthor
Kee-Tae Kim$^{6,7}$,
Paul F. Goldsmith$^{8}$,
A. Saha$^{2}$,
Dana Alina$^{9}$,
Chang Won Lee$^{6,7}$,
\newauthor
Saurabh Sharma$^{10}$,
Tie Liu$^{11,12}$,
Patricio Sanhueza$^{13}$,
Tapas Baug$^{14}$,
E. Sharma$^{2}$,
\newauthor
Fengwei Xu$^{15}$,
Ariful Hoque$^{14}$,
James O. Chibueze$^{16,17}$,
Dana Makarova$^{9}$, and
Wenyu Jiao$^{11}$\\
\thanks{Email: lokeshd@prl.res.in (LKD), naval@pku.edu.cn (NKB; Boya Fellow)}
$^{1}$Astronomy \& Astrophysics Division, Physical Research Laboratory, Navrangpura, Ahmedabad 380009, India.\\
$^{2}$Kavli Institute for Astronomy and Astrophysics, Peking University, 5 Yiheyuan Road, Haidian District, Beijing 100871, China.\\
$^{3}$National Astronomical Research Institute of Thailand (Public Organization), 260 Moo 4, T. Donkaew, A. Maerim, Chiangmai 50180, Thailand.\\
$^{4}$Faculty of Engineering, Gifu University, 1-1 Yanagido, Gifu 501-1193, Japan.\\
$^{5}$Indian Institute of Technology Gandhinagar Palaj, Gandhinagar 382355, India.\\
$^{6}$Korea Astronomy and Space Science Institute, 776 Daedeokdaero, Yuseong-gu, Daejeon 34055, Republic of Korea.\\
$^{7}$University of Science and Technology, Korea (UST), 217 Gajeong-ro, Yuseong-gu, Daejeon 34113, Republic of Korea.\\
$^{8}$Jet Propulsion Laboratory, California Institute of Technology, 4800 Oak Grove Drive, Pasadena, CA 91109, USA.\\
$^{9}$Physics Department, School of Sciences and Humanities, Nazarbayev University, Kabanbay batyr ave, 53, 010000 Astana, Kazakhstan.\\
$^{10}$Aryabhatta Research Institute of Observational Sciences, Manora Peak, Nainital 263002, India.\\
$^{11}$Shanghai Astronomical Observatory, Chinese Academy of Sciences, Shanghai 200030, People’s Republic of China.\\
$^{12}$State Key Laboratory of Radio Astronomy and Technology, Beijing 100101, People’s Republic of China.\\
$^{13}$Department of Astronomy, School of Science, The University of Tokyo, 7-3-1 Hongo, Bunkyo-ku, Tokyo 113-0033, Japan.\\
$^{14}$S. N. Bose National Centre for Basic Sciences, Block-JD, Sector-III, Salt Lake City, Kolkata 700106, India.\\
$^{15}$Max Planck Institute for Astronomy, K\"onigstuhl 17, 69117 Heidelberg, Germany.\\
$^{16}$UNISA Centre for Astrophysics and Space Sciences (UCASS), College of Science, Engineering and Technology, \\University of South Africa, Cnr Christian de Wet Rd and Pioneer Avenue, Florida Park, 1709, Roodepoort, South Africa.\\
$^{17}$Department of Physics and Astronomy, University of Nigeria, 1 University Road, Nsukka 410001, Nigeria.\\
}
\begin{document}

\date{ }

\pagerange{\pageref{firstpage}--\pageref{lastpage}} \pubyear{2026}

\maketitle

\label{firstpage}

\begin{abstract}
We present a multi-wavelength study of the massive protocluster G286.21+0.17 (G286) using \emph{JWST} near-infrared (NIR) imaging and ALMA H$^{13}$CO$^{+}$(1--0) observations. The \emph{JWST} images uncover a compact ($\sim$0.5 pc) hub-filament system (HFS), comprising a dense central hub connected by at least four converging filaments seen in absorption, along with multiple H$_2$ protostellar jets/outflows. The hub hosts the dense core G286c1. H$^{13}$CO$^{+}$ emission confirms this HFS over [$-$19.2, $-$16.4]~km~s$^{-1}$, while the \emph{JWST} images also trace prominent photodissociation regions surrounding H\,{\sc ii}~region~A, powered by a B-type star. The identified H$^{13}$CO$^{+}$ skeletons closely trace the major \emph{JWST} absorption structures and show steep velocity gradients indicative of inflow. 
The radial distribution of ALMA 1.3 mm continuum cores (ALMAGAL and Cheng et al. 2020) exhibits power-law trends toward the hub center. 
The core number density, surface density, and core mass follow $Y \propto r^{\alpha}$ with $\alpha_\rho$ $\sim$$-$2.8 to $-$1.1, $\alpha_\Sigma$ $\sim$$-$0.4 to $-$0.6, and $\alpha_M$ $\sim$$-$0.7, whereas the core diameter remains nearly constant. 
Together with filament mass accretion rates of 7.6--11$\times10^{-5}$ M$_\odot$ yr$^{-1}$, these results are consistent with competitive accretion, highlighting preferential core growth toward the hub.  
Increasing filament linewidths toward the hub imply gravity-driven inflow, but misaligned local velocity gradient and gravitational force directions along filaments point to a dynamically evolved system.
The HFS likely formed through large-scale gas-layer interactions aided by compression from the adjacent H\,{\sc ii} region. Overall, star formation in G286 appears to be regulated by filamentary accretion, competitive core growth in the hub, and stellar feedback. 
\end{abstract}

%
\section{Introduction} 
\label{sec:intro}
The processes governing the formation and early growth of massive OB stars ($>$ 8 $M_{\odot}$) and stellar clusters are still not fully understood. 
Growing observational evidence indicates that hub-filament systems \citep[HFSs;][]{myers09}, dense filamentary networks that channel material into a central hub are crucial sites for facilitating the rapid mass accumulation required for the formation of massive stars and stellar clusters \citep{Motte+2018,kumar20,padoan20,semadeni09,semadeni17,semadeni19,zhou22,yang23,maity23,maity25,bhadari25,dewangan25n}. This means that ongoing inflow of cold gas can progressively increase the mass of the hub, allowing the cluster to evolve into a more massive system. 
However, given the hierarchical nature of HFSs, which span spatial scales from $\sim$0.1--1 pc to $\gtrsim$10 pc \citep[e.g.,][]{kumar20,Bhadari2022}, the physical mechanisms governing mass inflow across these scales remain poorly understood. On small scales ($\lesssim$1 pc), gravity-driven accretion along filaments is widely 
recognized as a major process \citep[e.g.,][]{zhou22,Fengwei2023,bhadari25}. 
In contrast, on larger scales, mass inflow may be governed by alternative mechanisms, including pressure gradients induced by large-scale turbulent motions \citep[e.g., inertial inflow;][]{padoan20}, global gravitational collapse of the parent cloud, or the self-gravity of the filaments themselves. 

Furthermore, it is important to note that once a massive star or protocluster evolves, the signatures of HFSs weaken or disappear, at least for the smaller-scale structures within the hierarchy \citep{zhou22}. To interpret these scale-dependent inflow behaviors and their role in cluster formation, several theoretical scenarios have been proposed. These include the  global non-isotropic collapse (GNIC) scenario \citep{Tige+2017,Motte+2018}, the inertial inflow model \citep{padoan20}, and the Filaments to Clusters (F2C) model \citep{kumar20}. The GNIC picture unifies key aspects of competitive accretion \citep{bonnell01,bonnell04} with the global hierarchical collapse (GHC) paradigm \citep{semadeni09,semadeni17,semadeni19}. The combined high-resolution near-infared (NIR) observations from the James Webb Space Telescope (\emph{JWST}) and sub-millimeter observations from the Atacama Large Millimeter/submillimeter Array (ALMA) now offer an unparallel opportunity to probe the inner morphology and kinematics of star-forming regions with exceptional detail, enabling a critical assessment of the existing theoretical models. 
\begin{figure*}
\center
\includegraphics[width=\textwidth]{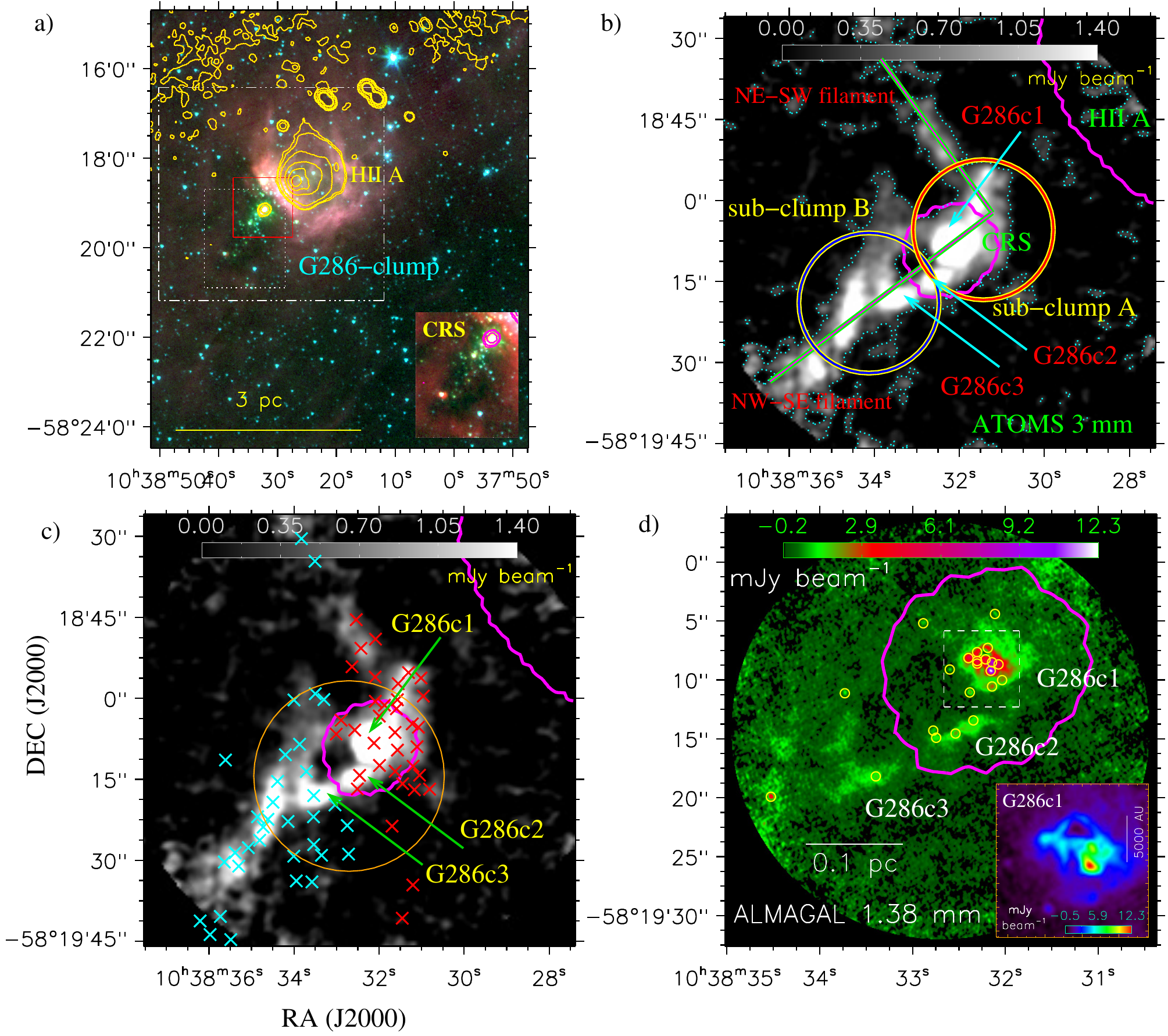}
\caption{a) Spitzer three-color composite map (8 $\mu$m in red, 4.5 $\mu$m in green, and 3.6 $\mu$m in blue) of the area hosting G286-clump and H\,{\sc ii}~region~A, overlaid with SMGPS 1.3 GHz continuum contours (yellow) at 0.075, 0.15, 0.55, 2, 3, and 3.6 mJy beam$^{-1}$. The bottom-right inset presents a zoomed-in view of G286-clump (see dotted box in Figure~\ref{fig1}a) using the same color scheme and contour levels. b) ATOMS 3 mm continuum map (see the solid box in Figure~\ref{fig1}a). The cyan dotted contour represents the 3 mm continuum emission at 0.23 mJy beam$^{-1}$, while the SMGPS 1.3 GHz continuum emission is shown by the magenta contour at 0.1 mJy beam$^{-1}$. Following \citet{zhou21}, sub-clumps A and B are indicated by large circles, and the locations of G286c1, G286c2, and G286c3 are marked with arrows. The NE--SW and NW--SE filaments are also labeled. c) ATOMS 3 mm continuum map overlaid with the SMGPS 1.3 GHz continuum contour (0.1 mJy beam$^{-1}$) and the positions of the 1.3 mm continuum dense cores (crosses) identified by \citet{cheng20b} (see also Figure~3 in \citealt{zhou21}). d) ALMAGAL 1.3 mm continuum map overlaid with the positions of the ALMAGAL 1.38 mm continuum cores (circles) associated with the ATLASGAL clump AG286.2092+0.1690 \citep{coletta25} and the SMGPS 1.3 GHz continuum contour (0.1 mJy beam$^{-1}$). The bottom-right inset presents a zoomed-in view of the G286c1 region enclosed by the dashed box. In panels ``a'' and ``b'', the compact radio source (CRS) and H\,{\sc ii} region A are labeled. A scale bar corresponding to a distance of 2.5 kpc is shown in panels ``a'' and ``d''.}
\label{fig1}
\end{figure*}

The focus of this study is the massive star-forming clump G286.21+0.17 or ATLASGAL clump AG286.2092+0.1690 (hereafter G286; also referred to as BYF73), located in the Carina spiral arm. Situated at a distance of 2.5 kpc \citep[e.g.,][]{barnes10,zucker20,zhou21}, G286 lies $\sim$12$'$ north of the rim of the 
Gum~31~bubble/H\,{\sc ii}~region \citep[diameter $\sim$15$'$;][]{barnes10}, which is ionized by the massive star cluster NGC 3324. A mass of $\sim$10$^{4}$ $M_{\odot}$ and a diameter of $\sim$0.9 pc were reported for the G286-clump \citep{barnes10}. Deep NIR observations from the Very Large Telescope presented by \citet{anderson17} revealed that a significant fraction of the young stellar objects (YSOs) in the embedded cluster still possess circumstellar disks, suggesting an extremely young evolutionary stage with an estimated age of $\sim$1 Myr. G286 is a very well studied star-forming site based on several multi-scale and multi-wavelength data sets including ALMA data \citep{cheng20a,cheng20b,cheng22,zhou21}. G286 is recognized as one of the most active and massive protocluster-forming regions \citep{faundez04,barnes10,ohlendorf13,anderson17,cheng20a,cheng20b,cheng22,zhou21}, hosting dense cores and large reservoirs of molecular gas embedded within the $\eta$~Car giant molecular cloud. 

Using ALMA Band 3 continuum and line data (beam size $\sim$2$''$), \citet{zhou21} identified two main filaments (NW-SE and NE-SW) along with two velocity-distinct sub-clumps (A and B). These authors also reported several compact dense cores (e.g., G286c1, G286c2, and G286c3) embedded in these filaments, which they suggested are actively participating in the formation of a young stellar cluster. Figure~\ref{fig1}a displays a three-color composite image of the region surrounding G286 constructed from {\it Spitzer} images overlaid with the radio continuum contours, while Figure~\ref{fig1}b presents the ALMA 3 mm continuum map of G286. The H\,{\sc ii}~region~A and the G286-clump are indicated in Figure~\ref{fig1}a, whereas the identified filaments, sub-clumps, and dense cores are marked in Figure~\ref{fig1}b. Velocity gradients detected across the filaments were attributed to large-scale compression flows influenced by nearby H\,{\sc ii} regions \citep[A and B; see Figure~1a in][]{zhou21}.
They observed double-peaked profiles in both optically thick and thin lines (e.g., HCO$^{+}$(1--0), H$^{13}$CO$^{+}$(1--0), and CCH (1--0)), attributing them to additional large-scale motions---such as the relative movement of the two sub-clumps or outflow activity---as well as to global infall within the clump. 
Based on single-dish spectral line profiles, the clump was previously interpreted as undergoing global gravitational infall \citep{barnes10}. 
The northern sub-clump A, which contains the known massive core G286c1, is characterized by strong dense-gas emission and ongoing massive star formation, whereas the southern sub-clump B is comparatively diffuse and less evolved. It has further been suggested that G286c1 is located at the intersection of the two filaments, whose collision produces the observed ``L''-shaped morphology \citep[see Figure~13 in][]{zhou21}. \citet{zhou21} did not identify G286 as an HFS \citep[see also][]{zhou22}, as clumps associated with evolved, spatially extended H\,{\sc ii} regions---seen as infrared bubbles in {\it Spitzer} 8 $\mu$m images---were excluded, as they are likely expanding and gravitationally unbound. 
This paper aims to investigate the physical processes driving massive-star and cluster formation in G286 and to re-assess its internal structure using \emph{JWST} NIR images together with ALMA 3 mm continuum and H$^{13}$CO$^{+}$(1--0) line data. This multi-wavelength approach provides a comprehensive characterization of the cloud morphology, gas distribution, and star formation activity in the region.

The structure of the paper is as follows: Section~\ref{sec:data} describes the observations, Section~\ref{sec:xxdata} presents the derived results, Section~\ref{sec:disc} provides the discussion, and Section~\ref{sec:conc} summarizes the main findings.
\section{Data and analysis} 
\label{sec:data}
In the direction of the G286-clump, we empolyed the ALMA 3 mm continuum and H$^{13}$CO$^{+}$(1--0) line 
data from the ATOMS survey \citep[Project ID: 2019.1.00685.S; PI: Tie Liu;][]{liu20a,liu20b}, which were also utilized by \citet{zhou21}. 
The observations were carried out with the ALMA 12-m and ACA 7-m arrays, and are primary-beam corrected \citep[see][for details]{zhou21}. The data have a synthesized beam size of $2\farcs62 \times 2\farcs40$ ($\sim0.032 \times 0.029$~pc at $2.5$~kpc), with a pixel scale of $0\farcs4$. 
The paper also used  science-ready \emph{JWST} Near-Infrared Camera \citep[NIRCam;][]{2005SPIE.5904....1R,2012SPIE.8442E..2NB} f356W, f410M, f444W+f405N (hereafter f405N), and f444W+f470N (hereafter f470N) images of G286.21+0.17 (Proposal ID: 3768; PI: Morten Andersen), which were retrieved from the Mikulski Archive for Space Telescopes (MAST; DOI: https://doi.org/10.17909/0d6v-3y92). Additional details on \emph{JWST} performance can be found in \citet{rigby2023}.

Additionally, the study incorporates {\it Spitzer} Galactic Legacy Infrared Mid-Plane Survey Extraordinaire \citep[GLIMPSE; resolution $\sim$2$''$;][]{benjamin03} data covering 3.6--8.0 $\mu$m, $^{13}$CO(1--0) line data from the Galactic Census of High and Medium-mass Protostars \citep[CHaMP; resolution $\sim$36$''$;][]{barnes11}, and 1.3 GHz continuum images from the South African Radio Astronomy Observatory (SARAO) MeerKAT Galactic Plane Survey \citep[SMGPS; resolution $\sim$8$''$;][]{Goedhart_2024}. 
In the direction of the ATLASGAL clump AG286.2092+0.1690 (G286), the positions and physical parameters of 
compact dust continuum cores/sources were collected from the recently published ALMAGAL catalog of 1.38 mm continuum detections \citep[][hereafter CO25]{coletta25}.  
The catalog is based on high-angular resolution ALMA 1.38 mm continuum observations (beam $\sim$0\rlap.{$''$}15--0\rlap.{$''$}3) acquired as part of the ALMAGAL survey \citep{molinari25}. The publicly available ALMAGAL 1.3 mm continuum map was also examined. For comparison, we used the 1.3 mm continuum core catalog of \citet[][hereafter CH20]{cheng20b}, derived from the combined ALMA 12 m and 7 m array continuum image with an angular resolution of $1\farcs62 \times 1\farcs41$. 
%
\section{Results}
\label{sec:xxdata}
\subsection{Multi-scale and multi-wavelength view of G286}
\label{sec:phy_env}
In this section, we examine the signatures of star formation activity across the broader G286 region.
\begin{figure}
\center
\includegraphics[width=0.9\linewidth]{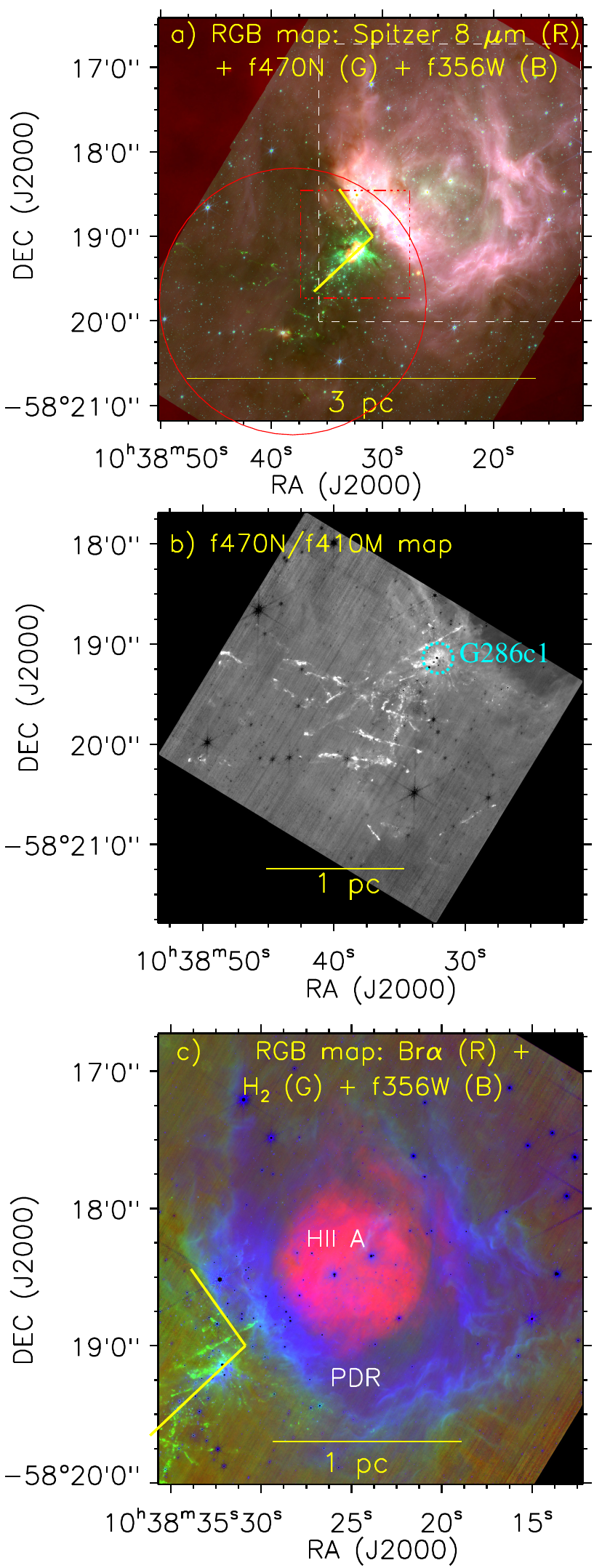}
\caption{a) Three-color composite map (8 $\mu$m (red)+ f470N (green) + f356W (blue)) made using {\it Spitzer} and \emph{JWST} NIR images (see the dot-dashed box in Figure~\ref{fig1}a). 
Most of the H$_{2}$ outflows seen within the encircled area in Figure~\ref{fg2}a are 
presented in Figure~\ref{fg2}b. b) \emph{JWST} f470N/f410M ratio map tracing multiple H$_2$ protostellar jets/outflows toward G286-clump. c) Three-color composite map generated using the f405N/f410M (Br$\alpha$; red), f470N/f410M (H$_{2}$; green), and \emph{JWST} f356W (blue) images (see the dashed box in Figure~\ref{fg2}a). In panels ``a'' and ``c'', the solid yellow lines highlight the NE-SW and NW-SE filaments, as indicated in Figure~\ref{fig1}b. In each panel, the scale bar is derived at a distance of 2.5 kpc.}
\label{fg2}
\end{figure}
\subsubsection{Detection of radio continuum emission toward G286}
\label{sec:c1env}
Figure~\ref{fig1}a displays a {\it Spitzer} three-color composite map constructed from the 8 $\mu$m (in red), 4.5 $\mu$m (in green), and 3.6 $\mu$m (in blue) images (resolution $\sim$2$''$), overlaid with the SMGPS 1.3 GHz continuum contours. In the northwest, the extended 8 $\mu$m emission outlines a shell-like structure, which appears to surround the 1.3 GHz continuum emission (i.e., H\,{\sc ii}~region~A). In this composite map, the intense 4.5 $\mu$m emission is seen extending across the entire G286-clump, and a compact radio source (CRS) is also detected toward the massive core G286c1 in the G286-clump. Using the same composite scheme along with the 1.3 GHz continuum emission contours, a zoomed-in view of the G286-clump is presented in the bottom-right inset of Figure~\ref{fig1}a. This inset highlights the area of enhanced 4.5 $\mu$m emission and marks the location of the CRS. 

In Figure~\ref{fig1}b, we display ATOMS 3 mm continuum emission map overlaid with its corresponding contours. The ALMA 3 mm continuum emission is predominantly concentrated toward the region exhibiting strong 4.5 $\mu$m emission. The previously known filaments, sub-clumps, and dense cores are also indicated in Figure~\ref{fig1}b. To highlight the spatial distribution of the ionized emission, the SMGPS 1.3 GHz continuum emission contour (in magenta) at 0.1 mJy beam$^{-1}$ is also shown in Figure~\ref{fig1}b. This overlay reveals two distinct regions of radio emission (i.e., CRS and H\,{\sc ii}~region~A), which are separated by a projected distance of $\sim$0.7 pc. Using the {\it clumpfind} IDL algorithm \citep{williams94}, we estimated the integrated flux densities of CRS and the H\,{\sc ii}~region~A from the SMGPS 1.3 GHz continuum map, obtaining values of 1.55 and 115.64 Jy, respectively. From these flux densities, the Lyman continuum photon rates (log $N_{\rm UV}$ in unit s$^{-1}$) are determined to be 44.88 and 46.75 for CRS and the H\,{\sc ii}~region~A, respectively. These values suggest that CRS and the H\,{\sc ii}~region~A are ionized by B2V and B0.5V-B0V type stars, respectively \citep{panagia73}. Further details of this analysis can be found in \citet{dewangan17a} \citep[see also][for the relevant equation]{matsakis76}.

In Figure~\ref{fig1}c, we show the positions of the CH20 cores (crosses) and the SMGPS 1.3 GHz continuum contour (0.1 mJy beam$^{-1}$) overlaid on the ATOMS 3 mm continuum map (see also Figure~3 in \citealt{zhou21}). These cores delineate the dense gas fragmentation within the G286.21+0.17 protocluster. Following \citet{zhou21}, the dense cores associated with sub-clump A are marked by red crosses, while those associated with sub-clump B are shown by cyan crosses. The observational coverage of the ALMAGAL 1.3 mm continuum map is indicated by the solid circle.
Figure~\ref{fig1}d presents the ALMAGAL 1.3 mm continuum map (angular resolution: $0\farcs39 \times 0\farcs34$), overlaid with the positions of 22 CO25 cores (circles) identified toward the ATLASGAL clump AG286.2092+0.1690 \citep{coletta25}, together with the SMGPS 1.3 GHz continuum contour (0.1 mJy beam$^{-1}$; beam size $\sim$8$''$). It allows to directly assess the correspondence between the continuum emission and the identified dense cores. The ALMAGAL observations improve the angular resolution from $\sim$$1\farcs5$ (\citealt{cheng20b}) to $\sim$$0\farcs3$, corresponding to a factor of $\sim$4 in linear resolution. As a result, several dense structures that appear as single continuum cores in the lower-resolution ALMA image are resolved into multiple compact dust continuum cores in the ALMAGAL data (see G286c1 and G286c2 in Figures~\ref{fig1}c and~\ref{fig1}d).

The area enclosed by the SMGPS 1.3 GHz continuum contour (extent $\sim$15$''$) contains 19 CO25 cores toward G286c1, while the remaining three cores lie outside the radio continuum emission contour. The physical diameters, masses, and average mass surface densities of these compact dust continuum cores are adopted from \citet{coletta25}; no independent core extraction or mass estimation is performed in this study. The cores exhibit a relatively narrow range of physical diameters (770--1630 AU). They span a wide range of masses (0.04--1.52 $M_{\odot}$) and mass surface densities (0.5--8.4 g cm$^{-2}$). Assuming a dust temperature of 41 K, \citet{coletta25} derived the core masses. Four cores have mass surface densities of $\Sigma \gtrsim 3$ g cm$^{-2}$, while twelve cores exceed $\Sigma \gtrsim 1$ g cm$^{-2}$ (not shown). A zoomed-in view of the region enclosed by the dashed box (i.e., G286c1) is presented in the bottom-right inset, highlighting the compact structures resolved by the ALMAGAL observations (see Figure~\ref{fig1}d). 

Although compact free-free emission from embedded ionized sources may contribute to the 1.3 mm continuum for a small number of cores located close to the CRS, assessing such contamination requires matched-resolution multi-frequency continuum observations and is beyond the scope of this work. Furthermore, the SMGPS 1.3 GHz observations (resolution $\sim$8$''$) and the ALMAGAL continuum data (resolution  $\sim$$0\farcs3$) probe very different spatial scales. Therefore, the integrated 1.3 GHz flux density of the compact radio source cannot be directly compared with the continuum flux densities of individual ALMAGAL cores. Nevertheless, the strong radio emission on $\sim$0.1 pc scales around G286c1 indicates that free-free contamination may be non-negligible for some of the nearby 1.3 mm sources. If present, such contamination would cause the dust-based continuum fluxes, and consequently the derived core masses, to be overestimated. The masses and other continuum-derived properties of sources embedded within the radio-bright and extended central emission should therefore be regarded as uncertain, and our interpretation of these individual sources is subject to this systematic uncertainty.

\begin{figure*}
\center
\includegraphics[width=\textwidth]{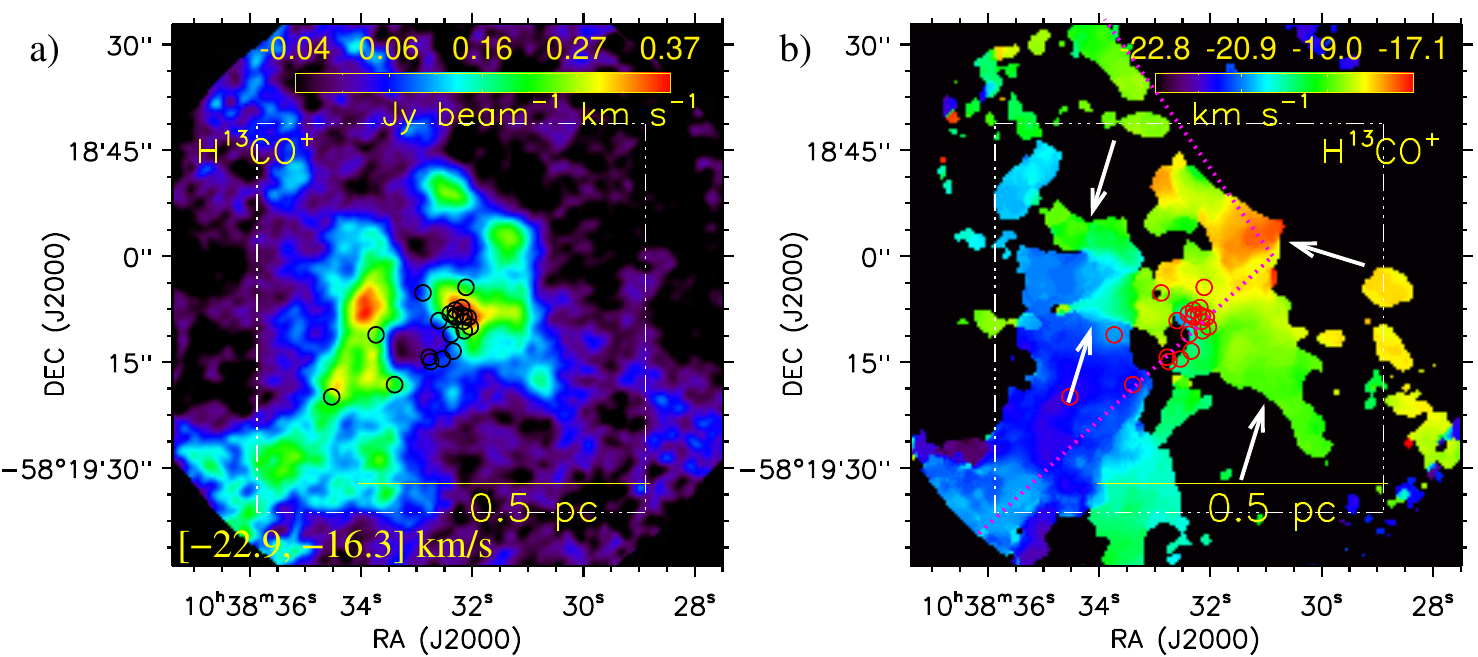}
\caption{a) ATOMS H$^{13}$CO$^{+}$(1--0) moment-0 map integrated over the velocity range of $-$22.9 to $-$16.3 km s$^{-1}$ (corresponding to the dot-dashed box shown in Figure~\ref{fig1}a). 
b) ATOMS H$^{13}$CO$^{+}$ moment-1 map. Arrows mark the candidate filaments forming an HFS morphology. Magenta dotted lines trace the L-shaped NE-SW and NW-SE filaments (see Figure~\ref{fig1}b). In each panel, the positions of the ALMAGAL 1.38 mm continuum cores associated with the ATLASGAL clump AG286.2092+0.1690 \citep[from][]{coletta25} are marked by open circles. A scale bar is shown assuming a distance of 2.5 kpc.}
\label{fig4}
\end{figure*}
\subsubsection{\emph{JWST} narrow band images: outflows and photodissociation regions}
\label{sec:c2env}
It is well established that the {\it Spitzer} 4.5 $\mu$m band includes the hydrogen recombination line (Br$\alpha$ at 4.05 $\mu$m) and strong molecular hydrogen (H$_{2}$) emission ($\nu$ = 0--0 $S$(9) at 4.693 $\mu$m), the latter of which is commonly excited by outflow shocks. Conversely, the {\it Spitzer} 3.6, 5.8, and 8.0 $\mu$m bands are primarily sensitive to polycyclic aromatic hydrocarbon (PAH) features at 3.3, 6.2, and 7.7 $\mu$m, which serve as excellent tracers of the photodissociation regions (PDRs) surrounding H\,{\sc ii} regions. With the narrow-band \emph{JWST} f405N and f470N images, we can further probe the specific zones exhibiting Br$\alpha$ and H$_{2}$ emissions, respectively.

A three-color composite map is presented in Figure~\ref{fg2}a, which is made of the {\it Spitzer} 8 $\mu$m (in red), \emph{JWST} f470N (in green), and \emph{JWST} f356W (in blue) images. This composite map highlights regions dominated by H$_{2}$ emission at 4.693 $\mu$m and additionally reveals a shell-like structure traced by the extended 8 $\mu$m and 3.565 $\mu$m emission.
\begin{figure}
\center
\includegraphics[width=1\linewidth]{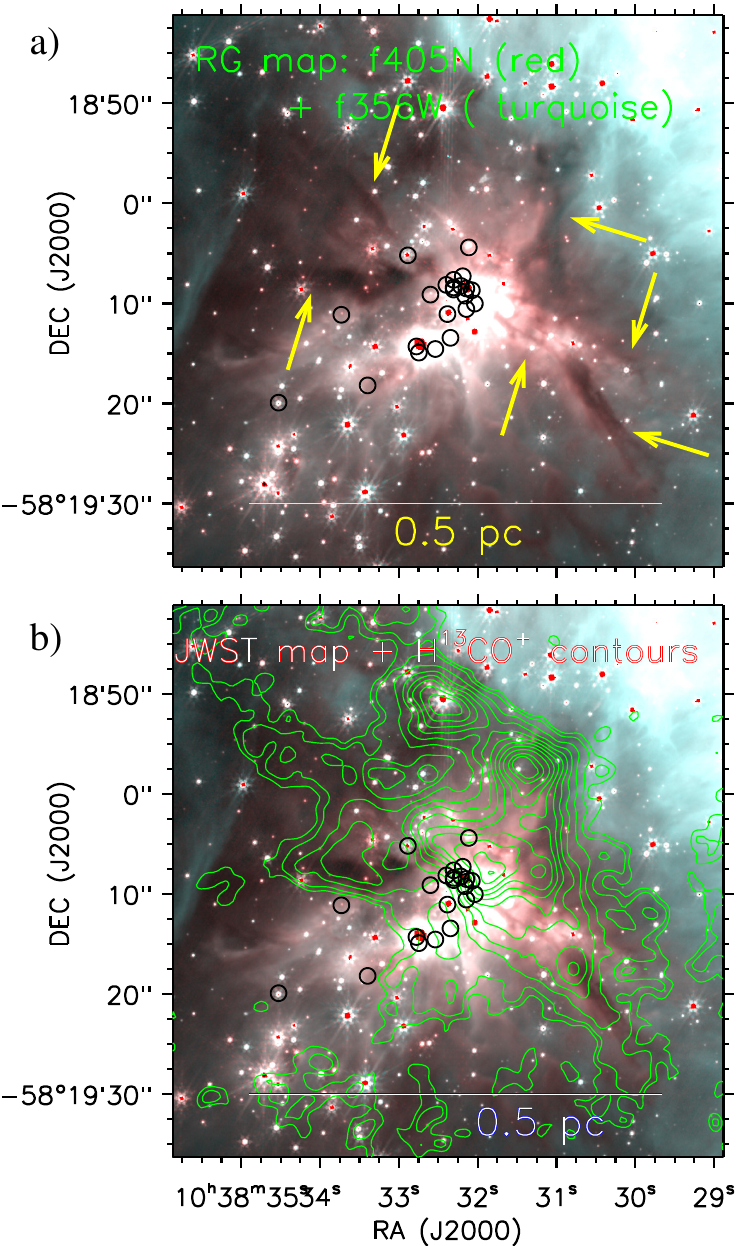}
\caption{a) Two-color composite \emph{JWST} image combining f405N (red) and f356W (turquoise), covering the region enclosed by the dot-dashed box in Figure~\ref{fig4}a. Filaments seen in absorption are indicated by arrows.
b) Same as Figure~\ref{fig5}a, overlaid with ATOMS H$^{13}$CO$^{+}$(1--0) integrated intensity contours (green) integrated over [$-$19.2, $-$16.4] km s$^{-1}$. Contour levels are 0.277 $\times$ [0.05, 0.1, 0.2, 0.3, 0.4, 0.5, 0.6, 0.7, 0.8, 0.9, 0.95] Jy beam$^{-1}$ km s$^{-1}$.}
\label{fig5}
\end{figure}

\begin{figure}
	\center
	\includegraphics[width=1\linewidth]{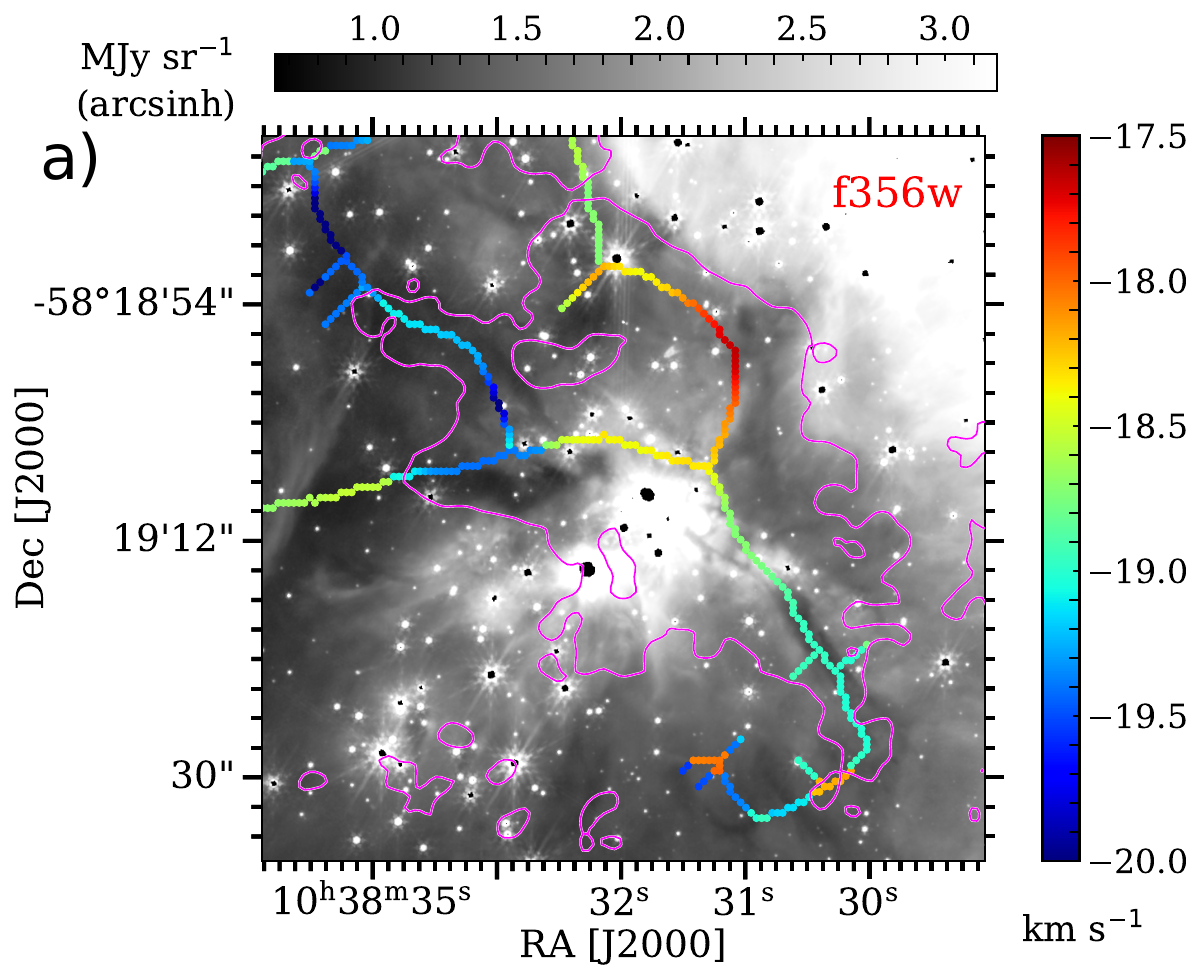}
	\includegraphics[width=0.9\linewidth]{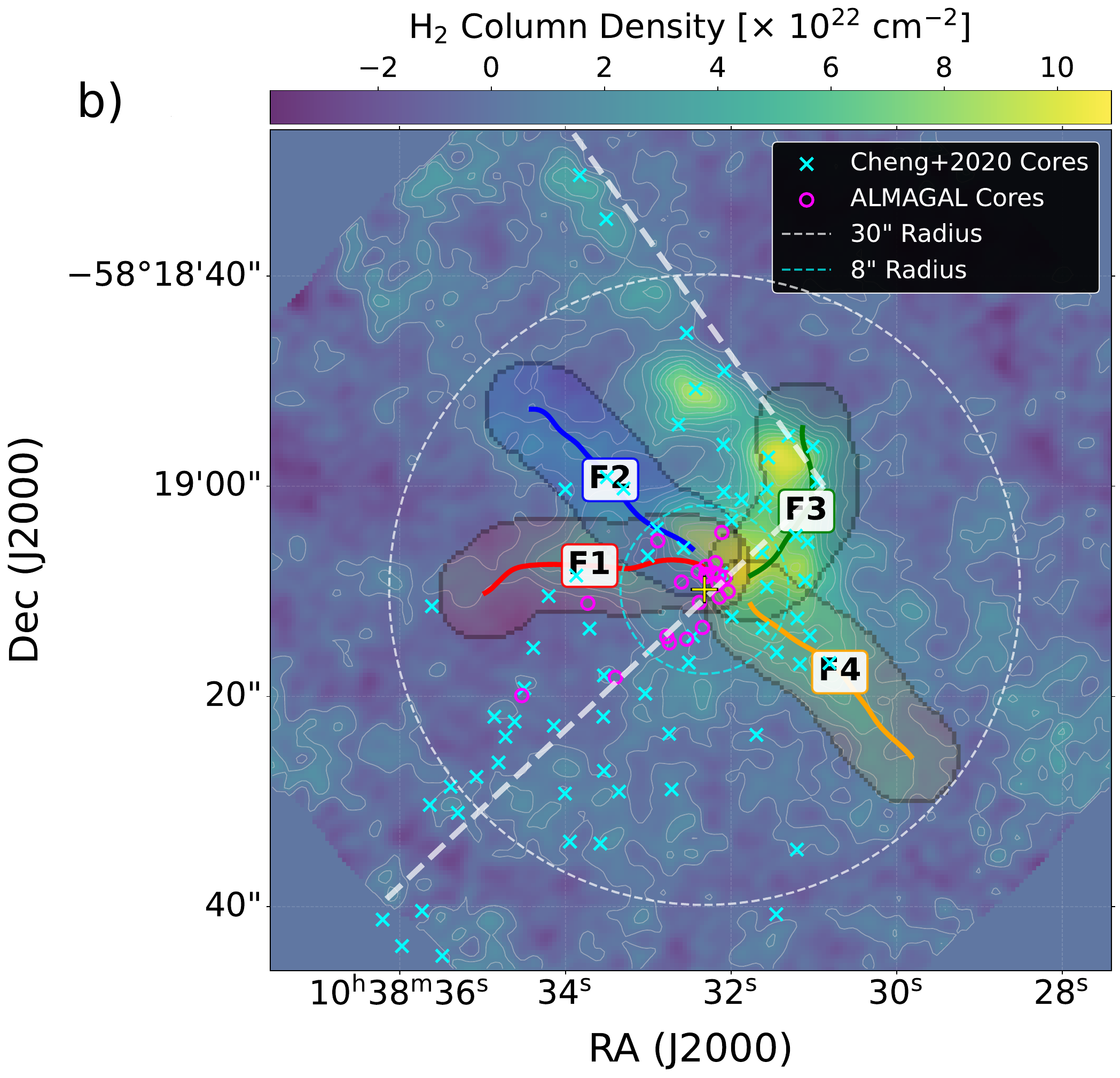}
	\caption{a) Overlay of the filamentary skeletons and the ATOMS H$^{13}$CO$^{+}$(1--0) integrated intensity contour (magenta; see also Figure~\ref{fig5}b) on the \emph{JWST} f356W image. These skeletons are identified using the \texttt{FilFinder2D} \citep{koch15} on the H$^{13}$CO$^{+}$ moment-0 map, with color indicating the H$^{13}$CO$^{+}$ velocity along the filaments (right color bar). b) H$_2$ column density map derived from H$^{13}$CO$^{+}$ emission, with contours from $(1$–$10)\times10^{22}$ cm$^{-2}$. Four filament spines (F1--F4) identified from the \emph{JWST} images, along with masks of width 0.1 pc, are 
	overlaid (see Figure~\ref{fig5}a). CH20 and CO25 core positions are displayed. Two circular regions of radii 8$''$ (marking a typical hub) and 30$''$ (covering HFS) centered on a $'+'$ symbol are displayed. White dashed lines trace the L-shaped NE-SW and NW-SE filaments (see Figure~\ref{fig1}b). }
\label{fgcolden}
\end{figure}
The H$_2$ emission detected in the \emph{JWST} maps can arise from either shock excitation or UV fluorescence. Shock-excited H$_2$ is typically produced by protostellar outflows or jets, where collisions in fast-moving gas heat the molecular material to high temperatures, resulting in bright and compact knot-like H$_2$ features. In contrast, fluorescently excited H$_2$ originates when UV photons from nearby massive stars irradiate molecular gas, generating more diffuse emission structures along the PDR interfaces. 
 To further examine the H$_2$ features, the \emph{JWST} ratio map (i.e., f470N/f410M) of f470N ($\lambda_{eff}$/$\Delta$$\lambda$: 4.708/0.051) and f410M ($\lambda_{eff}$/$\Delta$$\lambda$: 4.083/0.436) images is generated. The resulting map toward the G286-clump is presented  in Figure~\ref{fg2}b, where several H$_2$ protostellar jets/outflows are clearly seen. The location of the dense core G286c1 is also indicated in Figure~\ref{fg2}b (see the dotted circle).  The H$_{2}$ features are clearly resolved thanks to the high angular resolution of the \emph{JWST} infrared images. The bright, localized H$_2$ features observed in the f470N/f410M ratio map are therefore consistent with shock excitation associated with protostellar jet/outflow activities. 
To probe the 4.05 $\mu$m Br$\alpha$ emission, we have also generated the \emph{JWST} ratio map (i.e., f405N/f410M) of f405N ($\lambda_{eff}$/$\Delta$$\lambda$: 4.053/0.046) and f410M ($\lambda_{eff}$/$\Delta$$\lambda$: 4.083/0.436) images. 

Figure~\ref{fg2}c shows a three-color composite map constructed from the f405N/f410M (Br$\alpha$; red), f470N/f410M (H$_{2}$; green), and \emph{JWST} f356W (blue) images. In this RGB representation, the red channel traces Br$\alpha$ emission from ionized gas, 
the green channel highlights H$_{2}$ emission, and the blue channel corresponds to the F356W emission, which includes the 3.3 $\mu$m PAH feature. 
The morphology reveals that the Br$\alpha$ emission associated with H\,{\sc ii} region~A is enveloped by F356W emission, which is itself encircled by an exteneded layer of H$_2$ emission. This layered structure, comprising ionized gas, PAH-rich emission, and molecular hydrogen, is characteristic of a photodissociation region (PDR). The H$_2$ emission likely arises from fluorescent excitation at the molecular interface, where far-ultraviolet photons from the central B-type star interact with the surrounding molecular cloud.  
The embedded YSOs responsible for powering the H$_2$ outflows/jet-like features appear to be spatially offset from the PDRs. In the direction of the dense core G286c1, the detection of radio continuum emission without corresponding \emph{JWST} Br$\alpha$ emission is likely attributable to the combined effects of high extinction, limited Br$\alpha$ sensitivity, and the physical properties of the ionized gas.

\begin{figure*}
	\center
	\includegraphics[width=\textwidth]{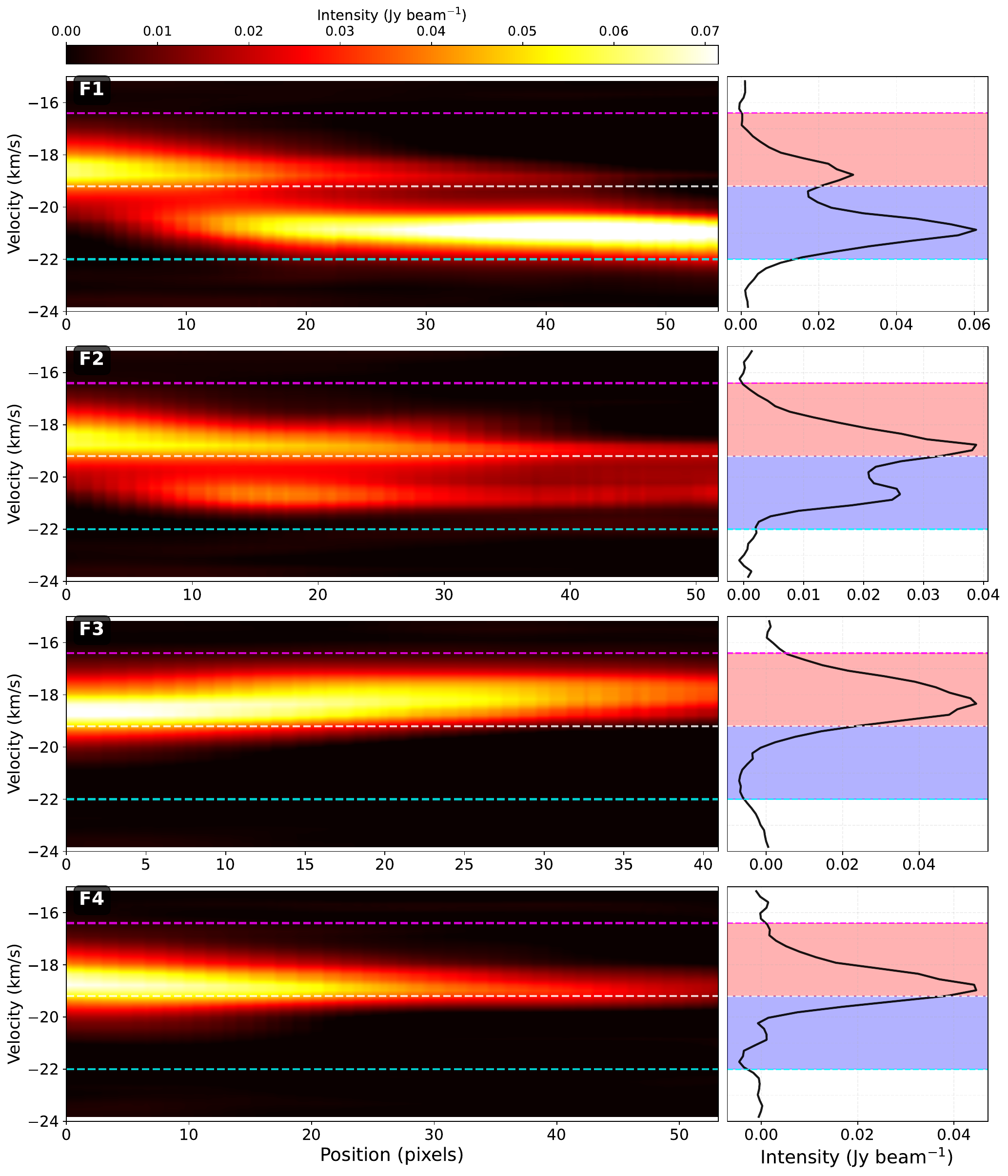}
	\caption{Position-velocity diagrams along the four identified filaments from head (hub) to tail (left column) averaged over a width of 0.1 pc (22 pixels) and averaged spectra within the filament regions (right column). The cyan, white, and magenta dashed lines mark $-$22, $-$19.2, and $-$16.4 km s$^{-1}$. The red and blue shaded regions in the spectra indicate the redshifted (HFS) and blueshifted cloud components, respectively.}
	\label{fgpv}
\end{figure*}
\begin{figure*}
\center
\includegraphics[width=\textwidth]{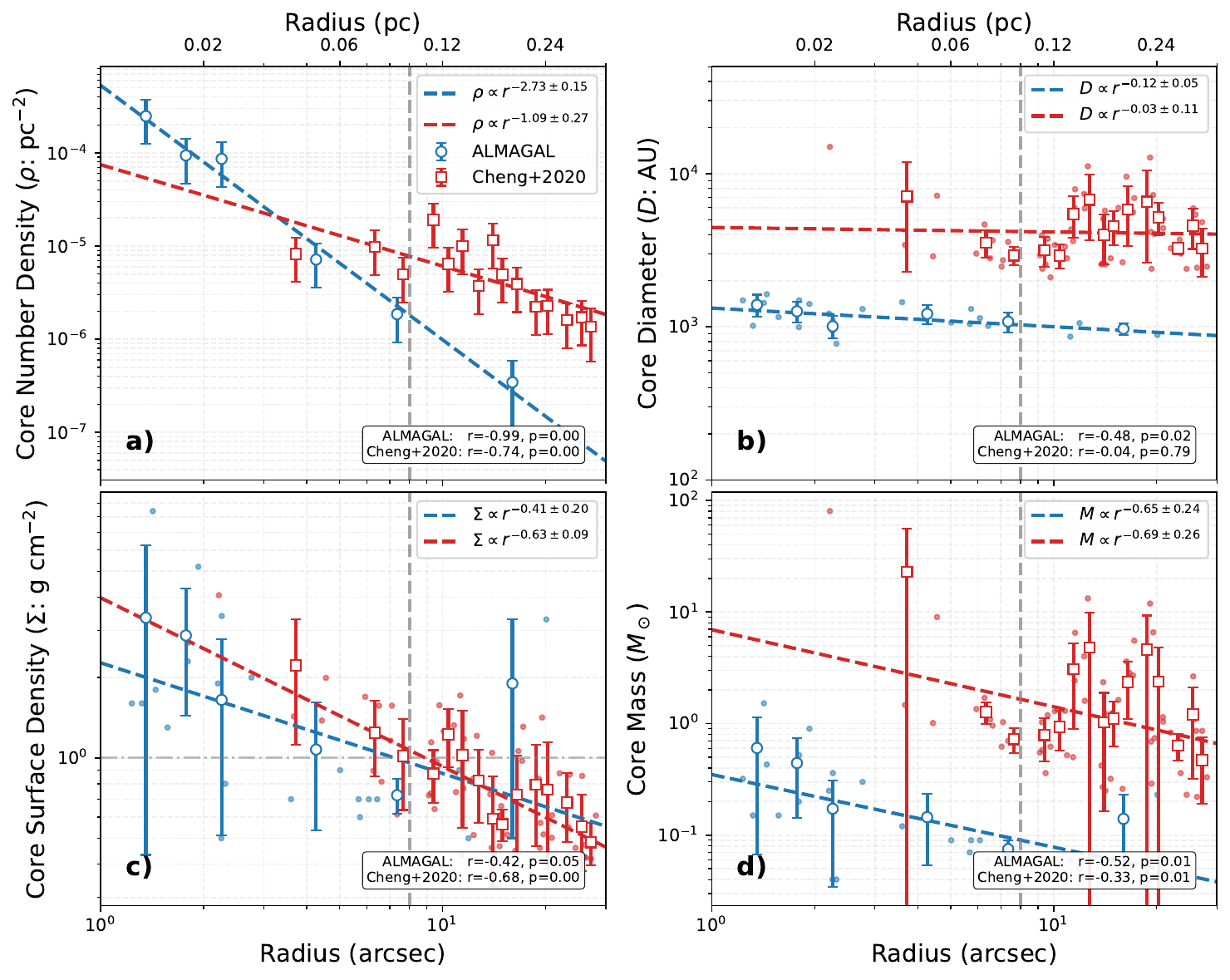}
\caption{Radial variation of core properties measured from the hub center: a) core number density ($\rho$, circles), 
b) core diameter ($D$, squares), c) core surface density ($\Sigma$, triangles), and d) core mass ($M_{\rm core}$, diamonds). 
The averaged value of the parameters for each bin is shown where the adaptive binning scheme is used considerig 4 cores per bin from $1''$ to $30''$ (bottom axis), with the corresponding physical  scale at $d= 2.5$ kpc shown on the top axis. 
Error bars represent $1\sigma$ standard deviations. 
A linear fit to the unbinned data point in panels b--c is shown in the log-log space, which is also labelled as power-law form of $Y \propto r^{\alpha}$. A vertical dashed line is mared at $r < 8''$ (see Figure~\ref{fig5}c). In panel ``c'', the horizontal dot-dashed line denotes the 1 g cm$^{-2}$ threshold for massive star formation \citep{Krumholz2008}.}
\label{cfig:coredistribution}
\end{figure*}
\subsection{Embedded hub-filament system in G286}
\label{linsec2}
To examine the internal structure and kinematics of the dense gas, we employ the ATOMS H$^{13}$CO$^+$(1--0) line data together with the \emph{JWST} images.  Previously, \citet{zhou21} analyzed the ATOMS HCO$^+$, H$^{13}$CO$^+$, CCH, and CS data over a wider velocity range. Their moment-0 and outflow-corrected moment-1 maps revealed distinct blueshifted and redshifted dense-gas components, an ordered large-scale velocity field, and multiple elongated dense structures rather than a single compact condensation (see their Figure~7). Among the ATOMS HCO$^+$, H$^{13}$CO$^+$, CCH, and CS data, we focus on the H$^{13}$CO$^+$(1--0) emission because it is generally optically thin and provides a reliable tracer of the intrinsic dense-gas kinematics with minimal contamination from self-absorption and molecular outflows. The integrated intensity (moment-0) and intensity-weighted velocity (moment-1) maps derived from the H$^{13}$CO$^+$(1--0) data are shown in Figures~\ref{fig4}a and \ref{fig4}b, respectively. 
The moment-0 map, integrated over the velocity range of $-22.9$ to $-16.3$ km s$^{-1}$, traces the distribution of dense molecular gas. The corresponding moment-1 map reveals a pronounced large-scale velocity gradient across the clump and indicates the presence of two morphologically distinct velocity components, which are also evident in the H$^{13}$CO$^{+}$(1--0) channel maps (see Figure~\ref{cfig:channelmap}). The first component is confined to velocities of $\sim-22$ to $-19.5$ km s$^{-1}$, whereas the second spans $\sim-19$ to $-16$ km s$^{-1}$ and exhibits a network of interconnected structures converging toward the central region. In the direction of this latter component, the candidate filaments are marked by arrows (see Figure~\ref{fig4}b). Both maps are overlaid with the positions of the 22 CO25 cores associated with the ATLASGAL clump AG286.2092+0.1690.  

Figure~\ref{fig5}a shows a two-color composite map produced using the \emph{JWST} f405N (red) and f356W (turquoise) images, which is overlaid with the positions of the CO25 cores. A compact central area containing these cores seems to be surrounded by at least four filamentary features seen in absorption (indicated by arrows in Figure~\ref{fig5}a), resembling an HFS candidate (i.e., G286-HFS). These filamentary absorption features are consistently visible in all the \emph{JWST} 
images (f356W, f410M, f405N, and f470N; see also Figure~\ref{fgx4} and Appendix~\ref{xxsec:jwstmap}).

In Figure~\ref{fig5}b, the ATOMS H$^{13}$CO$^{+}$(1--0) emission contours, integrated over 
the velocity range of $-$19.2 to $-$16.4 km s$^{-1}$, are overlaid on the \emph{JWST} color composite map. Interestingly, the molecular emission seems to follow the filamentary structures seen in the \emph{JWST} images, indicating that the candidate HFS in Figure~\ref{fig4}b is morphologically consistent with the \emph{JWST}-revealed HFS. To independently examine the molecular counterparts of the NIR absorption structures, we identify filamentary structures directly from the H$^{13}$CO$^{+}$ integrated-intensity map and derive their corresponding skeletons without imposing the \emph{JWST}-based filament geometry. 
We identified the filamentary structures in the H$^{13}$CO$^{+}$ moment-0 map ($[-16,-20]$ km s$^{-1}$) using the \texttt{FilFinder2D} algorithm \citep{koch15}. The map was flattened using the 96th percentile to remove large-scale background variations while preserving the extended filamentary emission, and the resulting mask was used to derive the filamentary skeletons. Figure~\ref{fgcolden}a shows these molecular skeletons overlaid on the \emph{JWST} F356W image, together with the H$^{13}$CO$^{+}$ integrated-intensity contour. To characterize the velocity gradients along these structures, we extracted spectra within a 3$''$ aperture centered on each skeleton pixel. A Gaussian profile was fitted to each spectrum, and the resulting centroid velocity was assigned to the corresponding skeleton pixel. In cases where two Gaussian components were required to fit the spectrum, the mean of the two centroid velocities was adopted. The centroid velocities are shown as a color scale along the filamentary skeletons in Figure~\ref{fgcolden}a. 
Several molecular structures spatially coincide with the prominent \emph{JWST} absorption features. This comparison provides an independent indication that the NIR-dark features are associated with coherent molecular structures. The velocity-coded molecular skeletons further reveal systematic velocity variations along individual structures and the convergence/overlap of multiple velocity components toward the central region. 

Figure~\ref{fgcolden}b presents the H$_2$ column density map derived from H$^{13}$CO$^{+}$ emission (see Appendix~\ref{sec:columndensity}), overlaid with the four filament spines identified by visual inspection of the \emph{JWST} F356W image. While the molecular-gas spines identified above were used to establish the morphological correspondence between the molecular structures and the NIR absorption features in the \emph{JWST} images, the filaments identified directly from the \emph{JWST} F356W image show a clearer radial alignment toward the central hub. We therefore prioritize these \emph{JWST}-identified spines in the analysis below and hereafter refer to them simply as the ``filaments''. A typical filament mask with a width of $\sim$0.1 pc, as commonly observed in nearby filamentary ISM structures, is also shown, together with the positions of the CH20 and CO25 cores. In addition, two circular regions with radii of $8^{\prime\prime}$ (representing the tentative hub) and $30^{\prime\prime}$ (encompassing the entire HFS) are indicated. The column-density map is primarily used to estimate the filament mass (Section~\ref{sec:massacr}). As noted earlier, the higher angular resolution of the ALMAGAL observations makes them more suitable than the ALMA continuum data of \citet{cheng20b} for investigating the spatial segregation of dense cores within the hub region (see the core distribution in Figure~\ref{fgcolden}b). The CO25 cores show a pronounced concentration toward the tentative hub, spatially coincident with the dense core G286c1. 

To investigate gas flows along the filaments, we constructed position-velocity (PV) diagrams for the four identified filaments. The PV diagrams were extracted along each filament spine using a mask width of 22 pixels ($\sim$0.1 pc), extending from the hub to the filament end (i.e., from head to tail), and are shown in Figure~\ref{fgpv} over the velocity range of $-24$ to $-15$ km s$^{-1}$. The corresponding velocity spectra, averaged within each filament mask, are displayed alongside the respective PV diagrams. The PV diagrams indicate that all four filaments are primarily associated with gas at velocities between $-19.2$ and $-16.4$ km s$^{-1}$. However, the PV diagrams of filaments F1 and F2 exhibit contamination from a blueshifted cloud component spanning velocities of $-22$ to $-19.2$ km s$^{-1}$. This additional component can be effectively separated by restricting the analysis to the velocity interval of $-19.2$ to $-16.4$ km s$^{-1}$, which corresponds to the HFS-associated cloud component.  
A more detailed view of the gas flow along the filaments can be obtained from the H$^{13}$CO$^{+}$(1--0) channel maps, presented in Figure~\ref{cfig:channelmap}, over the velocity range of $-22$ to $-16$ km\,s$^{-1}$ (see also Appendix~\ref{sec:columndensity}). The maps are overlaid with the visually identified spines of the four prominent filaments seen in the \emph{JWST} images, allowing a direct comparison between the gas emission and filamentary structure. The close spatial correspondence confirms the presence of a compact HFS, G286-HFS, with a size of $\lesssim$0.5 pc.
\subsubsection{Identification of hub and radial distribution of core properties}
\label{sec:hfsiden}
To quantify the radial distribution of core properties around the central hub, we first determined the hub center using a clustering analysis. The DBSCAN \citep[Density-Based Spatial Clustering of Applications with Noise;][]{Ester1996} algorithm was applied to the CO25 core positions, adopting a search radius of $5\arcsec$ and a minimum of three cores per cluster. This approach resulted in a largest DBSCAN cluster, containing 18 cores, whose center is defined as the mean position of its members. The derived coordinates ($\alpha = 10^{\mathrm{h}}38^{\mathrm{m}}32.33^{\mathrm{s}}, \quad \delta = -58^{\circ}19^{\prime}09.84^{\prime\prime}$) are taken as the hub center (see Figure~\ref{fgcolden}b). 
The remaining 4 cores are considered as noise, as they do not satisfy the adopted DBSCAN criteria. We note that the adopted hub centre coincides with the most massive core in both catalogs, corresponding to a mass of 80.24 $M_{\odot}$ in CH20 and 1.52 $M_{\odot}$ in CO25. The CH20 cores are likely to undergo further fragmentation at higher angular resolution. Nevertheless, for consistency, we examine the radial distributions of core properties using both core catalogs. The projected radial distance of each core from the hub center was computed to investigate the spatial variation of core properties. 
Figure~\ref{cfig:coredistribution} presents the radial distributions of the core number density ($\rho$, in cores pc$^{-2}$), core diameter ($D$, in AU), core mass ($M$, in $M_{\odot}$), and core surface density ($\Sigma$, in g cm$^{-2}$). 
To quantify the radial variation in $\rho$, we employed an adaptive binning scheme requiring a minimum of four cores per bin. The core number density was calculated as $\rho = N/A$, where $N$ is the number of cores in each bin (fixed at 4) and $A = \pi (r_{\mathrm{out}}^2 - r_{\mathrm{in}}^2)$ is the area of the corresponding annulus, with $r_{\mathrm{in}}$ and $r_{\mathrm{out}}$ denoting the inner and outer radii, respectively. The uncertainty in the core number density was estimated assuming Poisson statistics as $\sqrt{N}/A$.
For the remaining core properties (i.e., $D$, $M$, and $\Sigma$), Figure~\ref{cfig:coredistribution} shows both the individual core measurements and the binned averaged values, the latter represented by square symbols with $1\sigma$ standard errors.

To quantify the radial variation of the core properties, we fitted power-law relations of the form $Y \propto r^{\alpha}$ over a radial range extending to $30^{\prime\prime}$, encompassing the entire HFS structure shown in Figure~\ref{fgcolden}b. Both core catalogs exhibit a steep decline in number density with increasing radius, yielding power-law indices of $\alpha_\rho = -2.73 \pm 0.15$ for the CO25 sample and $\alpha_\rho = -1.09 \pm 0.15$ for the CH20 sample. The corresponding Pearson correlation coefficients are $-0.99$ and $-0.74$, respectively. Similar radial declines are found for the surface density, with $\alpha_\Sigma = -0.41 \pm 0.20$ (CO25) and $\alpha_\Sigma = -0.63 \pm 0.09$ (CH20), and for the core mass, with $\alpha_M = -0.65 \pm 0.24$ (CO25) and $\alpha_M = -0.69 \pm 0.26$ (CH20) (see Figure~\ref{cfig:coredistribution}).
We define the hub as the central region where the majority of the CO25 cores (19 out of 22) are concentrated and where all four filaments converge. This hub is delineated by the cyan circle of radius $8^{\prime\prime}$ ($\sim0.1$ pc) shown in Figure~\ref{fgcolden}b.

\subsubsection{Mass accretion rates along filaments in the G286-HFS}
\label{sec:massacr}
\begin{figure}
\center
\includegraphics[width=0.5\textwidth]{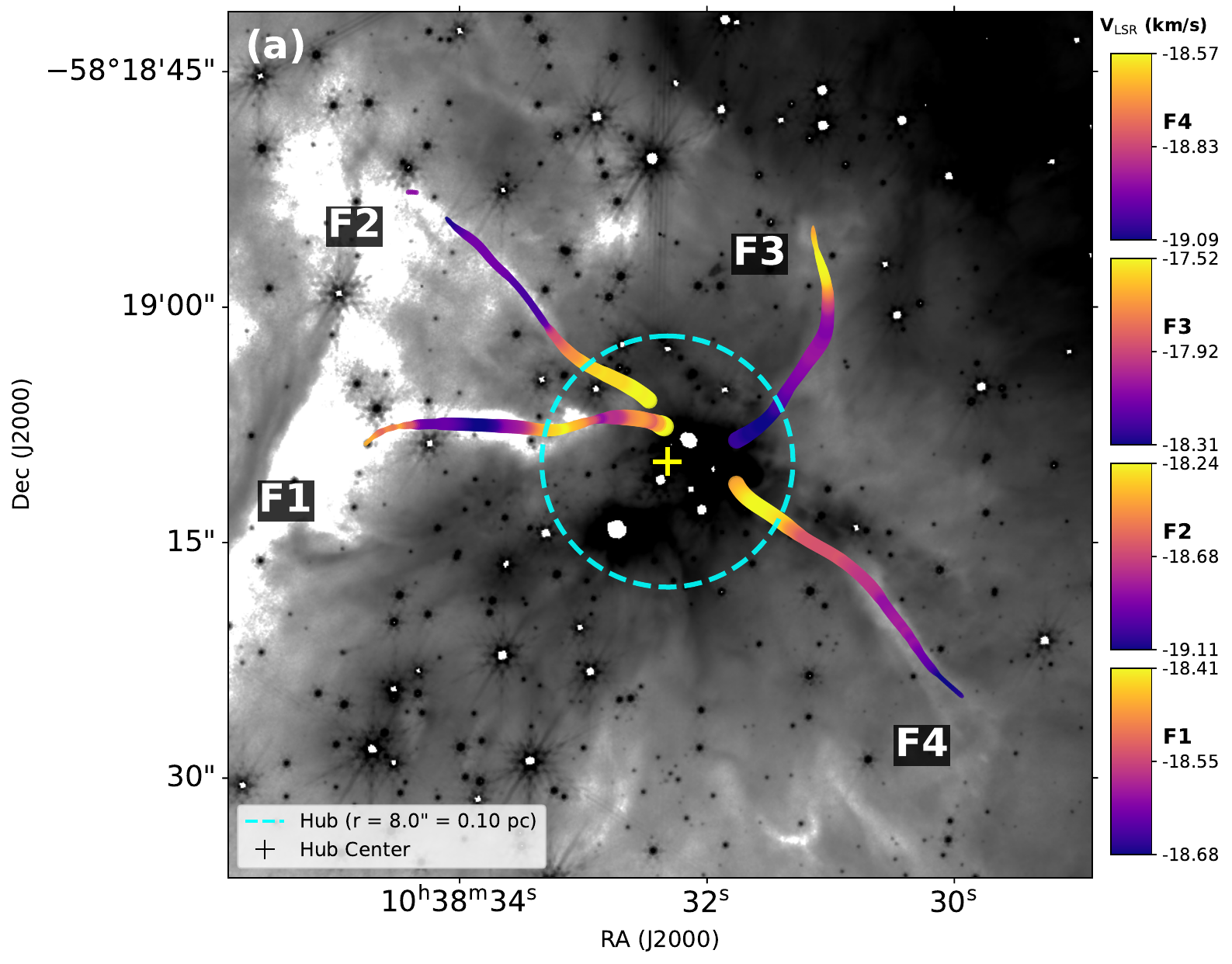}
\includegraphics[width=0.5\textwidth]{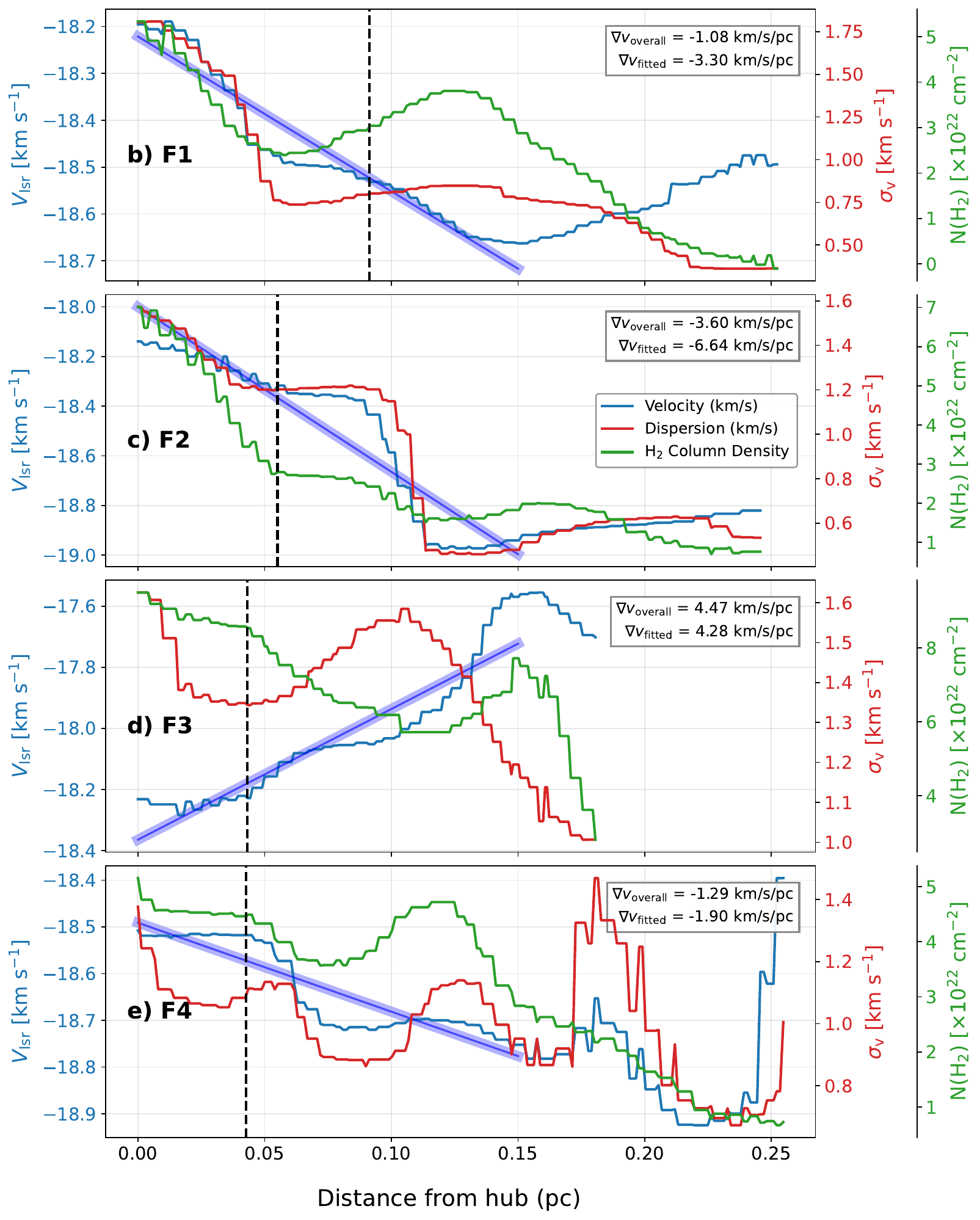}
\caption{a) Overlay of the selected filaments (F1--F4) on the inverted \emph{JWST} F356W image. 
Colors denote velocity, while spine thickness represents velocity dispersion. 
The color bars for the velocity of each filament are shown on the right. 
The circle and ``+'' symbol mark the hub region and its center, respectively (see Figure~\ref{fig5}c). b--e) Longitudinal profiles of velocity (blue), velocity dispersion (red), and H$_2$ column density (green) along the filaments, extracted using a circular aperture with a radius of 5 pixels. Vertical dashed lines mark the hub boundary, corresponding to the cyan circle in Figure~\ref{fig:hfs}a. The blue shaded strips represent the linear fit to the velocity gradient along filaments within range of 0--0.15 pc, with the global and fitted gradient value indicated in each panel. The x-axis shows the distance along the filaments measured outward from the hub.
} 
\label{fig:hfs}
\end{figure}
In Figure~\ref{fig:hfs}a, the filaments identified in the \emph{JWST} F356W image are highlighted, with the color scale representing the velocity along the filaments and the filament thickness indicating the linewidth dispersion. 
The velocity and linewidth were inferred from the H$^{13}$CO$^{+}$(1--0) moment-1 and moment-2 maps over the velocity range [$-$19.2, $-$16.4] km s$^{-1}$, respectively. The selection of this velocity range is discussed in Section~\ref{linsec2}. A visual inspection clearly reveals velocity gradient signatures along the filaments, with blueshifted-to-redshifted gas motion from the filament tail/outer-region toward the filament head/hub-region in filaments F1, F2, and F4, while the opposite trend is observed in filament F3. 
This may be due to the different projection of this filament compared to the others, supported by the fact that filament F3 lies at a far side based on its redshifted systemic velocity ($\sim-$18 km s$^{-1}$) compared to the others ($\sim-$18.7 km s$^{-1}$). Additionally, the influence of the H\,{\sc ii}~region shaping this filament cannot be neglected due to its close proximity. 
A more quantitative view is presented in Figures~\ref{fig:hfs}b--e, where longitudinal profiles of velocity, linewidth, and H$_2$ column density are shown for each filament, measured from the filament head (hub) toward the filament tail. These profiles are extracted using a circular aperture with a radius of 5 pixels. 
We have marked the hub boundary with vertical dashed lines for easy comparison, as the filaments were not drawn to a common hub center due to the lack of filamentary features toward the hub region in the {\it JWST} maps. Despite this, the filaments exhibit distinct behavior, with a clear enhancement in H$_2$ column density and velocity dispersion toward the hub region. 

The linewidth values for all filaments show a significant increase from the filament tail/outer-region to the filament head/hub-region, with typical gradients of 4--6 km s$^{-1}$ pc$^{-1}$. The presence of closely spaced velocity components may also contribute to enhanced linewidths, as is evident from the broader velocity emission toward the hub region shown in Figure~\ref{fgpv}. 
Filaments F2 and F4 exhibit similar velocity and linewidth profiles, possibly indicating more quiescent filaments hosting less star formation activity compared to F1 and F3. Local peaks in the H$_2$ column density profiles possibly indicate the presence of dense cores.

To quantify the velocity gradients along the filaments, we performed linear fits to the velocity profiles as a function of longitudinal distance from the hub outward (i.e., along the filament from head to tail). The absolute velocity gradients along the filaments are found to be in the range of 1.08--4.48 km s$^{-1}$ pc$^{-1}$ when considering the global gradients, and 1.90--6.64 km s$^{-1}$ pc$^{-1}$ when measured within 0--0.15 pc from the hub (see Figure~\ref{fig:hfs}). We estimate the mass accretion rate along the filaments using
	\begin{equation}
		\dot{M} = \frac{\Delta V_{\rm obs} M_{\rm fil}}{\tan(\alpha)},
	\end{equation}
where $\Delta V_{\rm obs}$ is the observed velocity gradient along the filament, $M_{\rm fil}$ is the filament mass (19.3--38.5 $M_\odot$), measured within the filament masks of 0.1 pc width (see Figure~\ref{fig5}c),
and $\alpha$ is the inclination angle of the filaments with respect to the plane of the sky \citep[assumed to be $45^\circ$; see][for further details.]{Kirk2013}. 
The filament masses were derived from the H$_2$ column density using the formula given in \citet{Bhadari2020}:
	\begin{equation}
M = \mu \, m_{\rm H} \, A_{\rm pix} \sum N_{\rm H_2}
	\end{equation}
where $\mu = 2.8$ is the mean molecular weight, $m_{\rm H} = 1.67 \times 10^{-24}$ g is the mass of a hydrogen atom, $A_{\rm pix}$ is the physical area of a pixel projected on sky-plane (in cm$^2$), and $\sum N_{\rm H_2}$ is the integrated H$_2$ column density over the filament mask.
The 0.1 pc width filament mask was defined based on common observations of filament widths in nearby star-forming filaments \citep{Arzoumanian2011}. Using the global velocity gradients of 1--4.5 km s$^{-1}$, the resulting mass accretion rates range from $2.13 \times 10^{-5}$ to $1.76 \times 10^{-4}$~$M_\odot~\mathrm{yr}^{-1}$, with a mean and median of $7.62 \times 10^{-5}$ and $5.37 \times 10^{-5}$~$M_\odot~\mathrm{yr}^{-1}$, respectively. In contrast, adopting the velocity gradients fitted over the inner 0--0.15 pc from the hub region (2--6.6 km s$^{-1}$) yields accretion rates of $4.84 \times 10^{-5}$ to $1.68 \times 10^{-4}$~$M_\odot~\mathrm{yr}^{-1}$, with a mean of $1.05 \times 10^{-4}$ and a median of $1.02 \times 10^{-4}$~$M_\odot~\mathrm{yr}^{-1}$. These values are, on average, $\sim$83\% higher than those derived using the global velocity gradients, indicating that the estimated accretion rates are sensitive to the local velocity field, which may be enhanced by the gravitational influence of the hub. We also note that the derived accretion rates are subject to several sources of uncertainty, primarily associated with the assumed H$^{13}$CO$^{+}$ abundance, excitation temperature used to estimate the column density (Section~\ref{sec:columndensity}), source distance, and the unknown inclination angle of the filaments. Varying the excitation temperature from 10 to 30 K changes the column density, and hence the filament mass and accretion rate, by a factor of $\sim$2. Similarly, adopting an H$^{13}$CO$^{+}$ abundance between $1.7\times10^{-11}$ \citep{Maruta2010} and $1.0\times10^{-10}$ \citep[e.g.,][]{Sanhueza2012} results in a variation of approximately a factor of 5. The combined effect of these uncertainties can alter the derived mass accretion rates by up to an order of magnitude.
The derived physical parameters of the filaments are listed in Table~\ref{tab:accretion_rates}. 
We further compared the 3 mm dust continuum derived column densities, which are typically 
	$\sim$1 $\times$ 10$^{22}~\mathrm{cm}^{-2}$ in dense regions and 
	$\sim$1.5$\text{--}3 \times 10^{21}~\mathrm{cm}^{-2}$ in diffuse regions. 
	The discrepancy between dust- and molecular-derived values is partly 
	attributable to the assumed H$^{13}$CO$^+$ abundance ratio, which varies 
	widely in the literature (e.g., \citealt{Sanhueza2012}). For the central 
	dense region in hub (radius $\sim$ 4$''$), the mean dust-derived N(H$_2$) is 
	$\sim$1 $\times$ 10$^{22}~\mathrm{cm}^{-2}$, compared to 
	$\sim$6.7 $\times$ 10$^{22}~\mathrm{cm}^{-2}$ from molecular data, a 
	systematic offset of $\sim$7. Because the filaments are relatively 
	diffuse, they are not prominently detected in dust continuum (except F3). 
	Toward the filamentary regions, the average N(H$_2$) from dust is 
	$\sim$1.5$\text{--}3 \times 10^{21}~\mathrm{cm}^{-2}$, while the 
	molecular-derived value is $\sim$2$\text{--}5 \times 10^{22}~\mathrm{cm}^{-2}$, 
	resulting in a mass difference of $\sim$10. Consequently, the accretion 
	rates are also affected by this factor.

\begin{table*}
	\centering
	\caption{Mass accretion rates derived from global and fitted velocity gradients.}
	\label{tab:accretion_rates}
	\begin{tabular}{lcccccc}
		\toprule
		Filament & $M_{\rm filament}$ & $|\nabla v_{\rm global}|$ & $\dot{M}_{\rm acc,global}$ & $|\nabla v_{\rm fitted}|$ & $\dot{M}_{\rm acc,fitted}$ \\
		& ($M_\odot$)    & (km s$^{-1}$ pc$^{-1}$) & ($M_\odot$ yr$^{-1}$)      & (km s$^{-1}$ pc$^{-1}$) & ($M_\odot$ yr$^{-1}$) \\
		\midrule
		F1 & 19.31 & 1.08 & $2.13e-05$ & 3.303 & $6.52e-05$ \\
		F2 & 20.25 & 3.60 & $7.45e-05$ & 6.637 & $1.37e-04$ \\
		F3 & 38.51 & 4.48& $1.76e-04$ & 4.277 & $1.68e-04$ \\
		F4 & 24.88 & 1.29 & $3.29e-05$ & 1.902 & $4.84e-05$ \\
		
		\midrule
		Mean & 25.74 & 2.61 & ${7.62e-05}$ & 4.030 & ${1.05e-04}$ \\
		Median & 22.56 & 2.45 & ${5.37e-05}$ & 3.790 & ${1.01e-04}$ \\
		Min & 19.31 & 1.08 & ${2.13e-05}$ & 1.902 & ${4.84e-05}$ \\
		Max & 38.51 & 4.48 & ${1.76e-04}$ & 6.637 & ${1.68e-04}$ \\
		\bottomrule
	\end{tabular}
\end{table*}

To assess whether the observed velocity gradients are driven by local gravity, we investigate the alignment between the on-sky velocity gradient and gravitational force vectors. In general, a preferential alignment between these vectors is indicative of gravity-driven gas flows in HFSs, as reported in previous studies (e.g., \citealt{Wang2022}; \citealt{bhadari25}).
To quantify the on-sky velocity structure, we derive the velocity gradient using second-order central differences:
	\begin{equation}
		G_{\xi}(i,j) = \frac{V_{\rm LSR}(i+1,j) - V_{\rm LSR}(i-1,j)}{2\Delta \xi}
	\end{equation}
	where $\xi$ represents either the right ascension ($\alpha$) or declination ($\delta$) coordinate, $(i,j)$ are the pixel indices, and $\Delta \xi$ is the pixel scale (i.e., 0\arcsec.4).
	
	The resultant velocity gradient magnitude and position angle are then computed as:
	\begin{align}
		V_{\rm g} &= \sqrt{G_{\alpha}^2 + G_{\delta}^2}, \\
		\theta_{V_{\rm g}} &= \frac{180^{\circ}}{\pi} \arctan\left(\frac{G_{\alpha}}{G_{\delta}}\right),
	\end{align}
	where $G_{\alpha}$ and $G_{\delta}$ are the gradients along the RA and Dec directions, respectively. 

We used the H$_{2}$ column density map to compute the gravitational 
force vector ($F_{G}$) map. The gravitational force at pixel 
$i$ is estimated as the vector sum of contributions from all pixels 
within the selected region:
	
\begin{equation}
F_{G,i} = k \, I_i \sum_{j=1}^{N} \frac{I_j}{r_{ij}^2} 
\hat{r}_{ij},
\label{eq:grav_force}
\end{equation}
	
\noindent where $I_i$ and $I_j$ are the pixel intensities at positions 
$i$ and $j$, $r_{ij}$ is the projected distance between them, 
$\hat{r}_{ij}$ is the unit vector directed from $j$ toward 
$i$, and $N$ is the total number of pixels. The constant $k$ 
encapsulates the column density-to-mass conversion and the 
gravitational constant. Since we are interested only in the 
direction of $F_{G,i}$, we set $k = 1$. 	

\begin{figure*}
\center
\includegraphics[width=\textwidth]{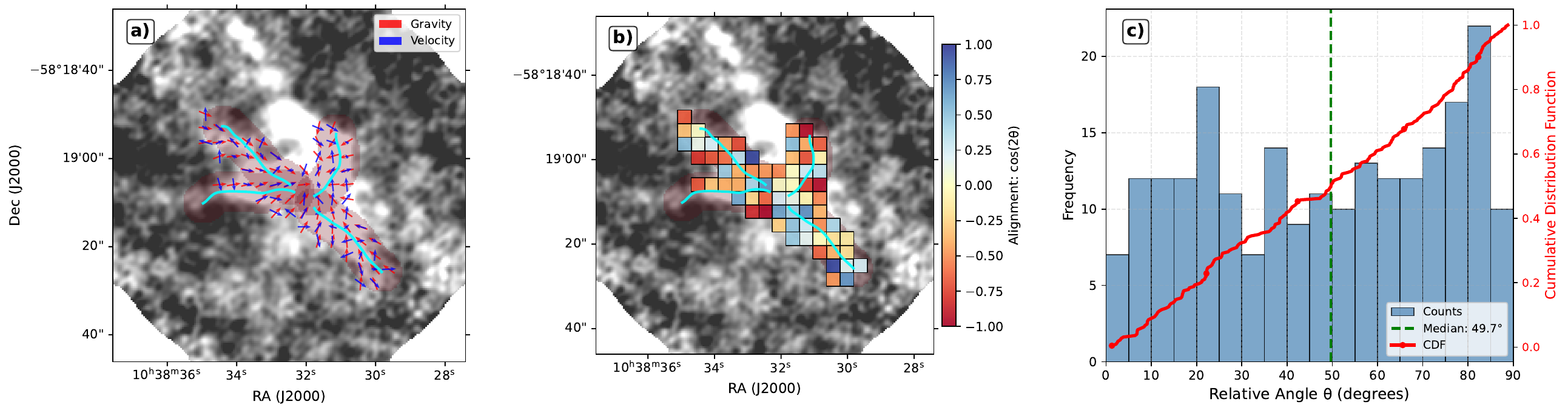}
\caption{Distribution of relative angles between the on-sky velocity gradient and gravity vectors, shown as: a) vector overlay, and b) the alignment measure (AM) parameter, where AM = 1 corresponds to perfectly aligned vectors and AM = $-$1 indicates perpendicular vectors. The background corresponds to the H$^{13}$CO$^{+}$ moment-0 map, with filament spines marked by cyan curves. c) Histogram of the relative angle distribution, overlaid with its cumulative distribution function. The vectors and AM values are averaged over 8-pixel cells to minimize overlap.}
\label{fig:vel_gravity}
\end{figure*}

Figure~\ref{fig:vel_gravity}a shows the distribution of on-sky velocity gradient and gravitational force vectors overlaid on the H$^{13}$CO$^{+}$ moment-0 map of the HFS region. For visual clarity, the filament spines are also overlaid. We find no clear alignment between the velocity gradient and the gravitational force vectors. 
Figure~\ref{fig:vel_gravity}b quantifies this behavior by presenting the distribution of the alignment measure $\mathrm{AM}=\cos(2\theta)$ between individual vector pairs, where $\mathrm{AM}=1$ indicates parallel alignment and $\mathrm{AM}=-1$ indicates perpendicular alignment. The histogram distribution of relative angles is shown in Figure~\ref{fig:vel_gravity}c. In most cases, the relative orientations exhibit no preferred alignment and instead appear consistent with a random distribution. To avoid overlap and ensure consistent spatial sampling, the vectors and AM values are gridded to $8\times8$ pixel cells. 
\section{Discussion}
\label{sec:disc}
The molecular cloud associated with G286 is traced by CHaMP $^{13}$CO(1--0) emission in the velocity range of [$-$23.2, $-$16.4] km s$^{-1}$ (see Appendix~\ref{Asec:coem} and Figure~\ref{fig:apx1}a). 
The cloud hosts the G286-clump, which exhibits prominent filamentary structures in absorption along with H$_2$ jets/outflows in the \emph{JWST} NIR images, as well as a shell-like morphology associated with H\,{\sc ii}~region~A (see Figures~\ref{fg2} and ~\ref{fig5}).

Observations with {\it Spitzer} and {\it Herschel} along with radio continuum maps have demonstrated that MIR shells or bubbles associated with H\,{\sc ii} regions, filaments, and young stellar clusters commonly coexist in massive star-forming regions, highlighting the interplay of 
multiple physical processes such as massive-star feedback and filamentary accretion flows \citep{Churchwell2006,Deharveng2010,andre14}. The advent of \emph{JWST} has further advanced this field by providing high-resolution narrow- and broad-band NIR observations that reveal these structures in unprecedented detail \citep[e.g.,][]{Dewangan2023,Dewangan2024a,Dewangan2024b,jadhav25}. 

Using ALMA observations ($\sim$1$''$ resolution) of G286, \citet{cheng20b} revealed highly filamentary C$^{18}$O emission concentrated toward the central protocluster region and associated with the dust continuum. They found two spatially distinct velocity groups of dense cores, suggesting ongoing infall and a dynamically unrelaxed system. 
\citet{barnes23} characterized the G286-cloud’s magnetic fields, gas dynamics, and energetics, and reported a powerful, dynamically young ($<$1 Myr) CO outflow from the massive protostar MIR~2 toward G286c1.

Our combined analysis of \emph{JWST} NIR images and ALMA H$^{13}$CO$^{+}$(1--0) line data reveals a new and more comprehensive picture of the G286 region. These observations provide strong evidence for a compact/small-scale HFS with a size $\sim$0.5~pc, in which a central dense hub is connected to at least least four converging filaments (F1--F4). In particular, the inclusion of \emph{JWST} data reveals extinction filaments and their configuration that were not identified in earlier studies \citep[e.g.,][]{cheng20b,zhou21}.  
We note that the ALMA C$^{18}$O observations presented by \citet{cheng20b}, covering a larger spatial extent, reveal filamentary structures spanning $\sim$2--3 pc that converge toward the L-shaped feature and appear to merge with the NW--SE filamentary ridge or the larger hub (see their Figure 5). This suggests that the HFS identified in this work ($<$1 pc) represents a compact, dense substructure embedded within the larger-scale (2--3 pc) filamentary network reported by \citet{cheng20b}. These observations support a hierarchical picture of mass assembly, in which parsec-scale filaments funnel material into a dense ridge, while the densest portion of the ridge further fragments into a compact HFS  with a characteristic scale of $\sim$0.5 pc, as revealed by the \emph{JWST} and ALMA observations. This interpretation is consistent with a multi-scale mass assembly scenario for HFSs \citep{zhou21}. 

The previously known massive core G286c1 as well as CO25 cores are predominantly concentrated toward the central hub of this HFS, which is traced in the velocity range of $[-19.2,-16.4]$~km~s$^{-1}$. In addition to the HFS, multiple H$_2$ protostellar jets/outflows are also evident in the \emph{JWST} images (see Figures~\ref{fg2}a and~\ref{fg2}b), and may be associated with other dense cores in the G286-clump. 
The central hub of the HFS, which hosts the dense core G286c1, is also associated with the SMGPS 1.3 GHz continuum emission (i.e., CRS). 
The velocity component ($[-19.2,-16.4]$~km~s$^{-1}$) appears morphologically associated with the filamentary network identified in the \emph{JWST} images (see Figure~\ref{cfig:channelmap}), reinforcing the physical association between the dense gas and the embedded stellar population. 
Several of the ALMAGAL cores have surface densities $\Sigma \gtrsim 3$ g cm$^{-2}$ (see Section~\ref{linsec2}). The coexistence of these cores with a large population of very low-mass, low-$\Sigma$ cores suggests a hierarchical and non-coeval core formation scenario, consistent with the competitive accretion model.

Figure~\ref{cfig:coredistribution} illustrates this picture, showing that the gravitational center of the cluster (i.e., the hub center) hosts a large number of cores, including the most massive and highest surface-density objects. Both the core mass and surface density decrease with increasing distance from the hub center, accompanied by a decline in the core population. In contrast, the core sizes remain approximately constant within the hub. 
We define the hub boundary at $r\simeq0.1$~pc, based on the concentration of dense cores and, primarily, the spatial convergence of the filamentary structures toward G286c1. The radial distributions of core mass, surface density, and core number density show an overall increase toward the central hub region, although with substantial scatter. We caution that these trends should not be regarded as a direct quantitative test of competitive-accretion models, since the observed power-law slopes are affected by temperature assumptions, spatial resolution, sensitivity, and projection effects. Nevertheless, the enhanced concentration of mass and the increasing number of sub-core structures toward G286c1 are qualitatively consistent with a scenario in which gravitational focusing promotes preferential growth of dense structures toward the hub. The CO25 and CH20 catalogs probe different spatial scales, and a single CH20 core can contain multiple substructures; therefore, their radial slopes do not represent identical physical quantities. Their broadly similar radial trends instead suggest that the central mass concentration persists across spatial scales.

In the G286 HFS, the estimated mass accretion rates along the filaments (F1--F4) range from 7.6--11$\times$10$^{-5} ~M_\odot~\mathrm{yr}^{-1}$ (see Section~\ref{sec:massacr}).  
\citet{padoan20} showed that mass-flow rates toward prestellar cores scale nearly linearly with spatial size, increasing from $\sim10^{-5}$ M$_\odot$ yr$^{-1}$ at $\sim0.1$ pc to $\sim10^{-4}$ $M_\odot$ yr$^{-1}$ at $\sim1$ pc, consistent with observational measurements \citep{Yuan2018,Sanhueza2021,Sanhueza2025,bhadari25,Morii2025,Wells2025,Gupta2026}. The inflow rates derived in this work fall within a similar range, also comparable to those reported for feedback-dominated regions \citep[e.g.,][]{Wells2025}. Nevertheless, these estimates should be treated with caution, since the filament masses inferred from molecular line emission are subject to systematic uncertainties arising from assumptions of optically thin emission. Additionally, projection effects and uncertainties in filament geometry can bias the inferred values; for example, inclination angles between $30^\circ$ and $60^\circ$ may lead to uncertainties of up to a factor of $\sim2$ in the estimated accretion rates.

Furthermore, the linewidths measured along these filaments decrease markedly from the head toward the tail, with typical gradients of 4--6 km s$^{-1}$ pc$^{-1}$ (see Figure~\ref{fig:hfs}). This behavior likely reflects gravity-driven turbulence and core formation, as suggested by \citet{Peretto2014}. However, we find no evidence for a preferential parallel alignment between the velocity gradients and the gravitational force vectors along the filaments  (see Figure~\ref{fig:vel_gravity}c), implying that coherent gravity-driven inflows are not clearly traced in the present kinematic state of the system. This lack of alignment may be a consequence of the relatively evolved nature of the G286-HFS, where stellar feedback from the forming star cluster may have weakened or erased the signatures of coherent gas inflow. 
This behavior contrasts with the G11P1-HFS studied by \citet{bhadari25}, where a preferential alignment between gravitational force and velocity gradient vectors was reported, 
indicative of a younger system prior to the onset of significant feedback.  
Additionally, the lack of a coherent velocity gradient and gravitational signature may suggest that the region could be shaped by random turbulent compression rather than gravitational inflow \citep{Federrath2012,Padoan2011}. 

As mentioned, the \emph{JWST} NIR observations have revealed 
widespread H$_2$ jets/outflows and PDRs around H\,{\sc ii}~region~A. The combination of the 4.69 $\mu$m narrow-band observations with Br$\alpha$ emission traced by the f405N filter enables an efficient separation of shock-excited emission associated with outflows and stellar feedback from fluorescently excited features. 
Using the SOFIA spectral line and continuum data, \citet{pitts18} also studied the PDR around the H\,{\sc ii}~region in G286. The driving sources of the H$_2$ outflows appear to be located away from the boundary layers between ionized gas and the neutral/molecular PDR (see Figure~\ref{fg2}c), suggesting that ongoing star formation is not confined to the immediate feedback dominated interfaces.  
 
The observed HFS in G286 may reflect signatures of the onset of the GNIC scenario \citep{Tige+2017,Motte+2018}. In other words, the observed HFS suggests ongoing mass accretion along filaments, while the presence of the PDR around the H\,{\sc ii}~region~A indicates the influence of stellar feedback. The coexistence of these two signatures suggests that the clump, associated with intense star formation activities, is evolving under the combined influence of ongoing accretion and stellar feedback, rather than being dominated by either process alone.  

\citet{zhou21} reported a collision between two orthogonal filaments (NE-SW and NW-SE), forming an L-shaped structure with the interaction junction coincident with the protocluster associated with subclump~A. 
Our results support the interpretation of \citet{zhou21} that H\,{\sc ii}~region~A interacts with
the NE-SW filament. The higher-resolution JWST images and dense-gas kinematics
further show that this filament is linked to the central hub-filament network. Supporting evidence for such expansion is provided by the position-velocity diagram of CHaMP $^{13}$CO(1--0) emission, which shows an arc-like structure (Figure~\ref{fig:apx1} and also appendix~\ref{Asec:coem}). Such arc- or C-shaped features are well-known kinematic signatures of expanding H\,{\sc ii} regions \citep[e.g.,][]{Arce2011}. The impact of massive-star feedback is further supported by pressure estimates \citep[e.g.,][]{dewangan17a}. At a projected distance of 0.7 pc, corresponding to the separation between the CRS and H\,{\sc ii} region~A (see the magenta contour in Figure~\ref{fig1}b), the combined pressure from the H\,{\sc ii} region, radiation, and stellar winds 
($P_{\rm total} = P_{\rm HII} + P_{\rm rad} + P_{\rm wind}$) driven by a massive B-type star is estimated to be 
$\gtrsim 10^{-10}$ dynes cm$^{-2}$ \citep[e.g.,][]{dewangan17a,sharma24}. 
This pressure exceeds that of a typical cold molecular cloud 
($P_{\rm MC} \sim$$10^{-11}$--$10^{-12}$ dynes cm$^{-2}$), assuming a temperature of $\sim$20~K and particle densities of 
$\sim$$10^{3}$--$10^{4}$ cm$^{-3}$ \citep[see Table~7.3 of][]{dyson80}. 

The origin of the HFS in G286 likely involves multiple processes acting in concert. One possibility is that interactions between gas layers with different velocities could produce the observed HFS morphology and could trigger protocluster formation (see velocity components at $-18.76$ and $-20.87$ km s$^{-1}$ in the bottom-right panel of Figure~\ref{cfig:channelmap} and also Appendix~\ref{sec:columndensity}). Such collisions between gas layers have been proposed as a viable mechanism for HFS formation in several recent studies \citep[][and references therein]{Maity2024ccc,Georgatos2026arXiv}. Alternatively, or additionally, feedback from H\,{\sc ii}~region~A may have enhanced the accumulation of dense gas along the filament NW-SE, thereby promoting interactions among multiple gas layers. Given the close proximity of H\,{\sc ii}~region~A and its powering by a B-type star, this feedback is likely to have played an important role in the formation of the 
protocluster as clearly seen in the \emph{JWST} color-composite map (Figure~\ref{fg2}c).

Collectively, our results imply that massive star and cluster formation in G286 arises from the joint action of ongoing accretion, competitive core growth in the hub, and stellar feedback.
\section{Summary and conclusion}
\label{sec:conc}
To develop a comprehensive picture of the physical processes operating in the massive star- and cluster-forming region G286, we carried out a multi-wavelength, multi-scale analysis by combining observations from various facilities and surveys, including \emph{JWST}, {\it Spitzer}-GLIMPSE, CHaMP, ALMA/ATOMS, ALMA/ALMAGAL, and SMGPS. The major outcomes of this work are outlined as follows:
\begin{itemize}
\item \emph{JWST} NIR images reveal a compact ($\sim$0.5 pc) hub-filament system (G286 HFS), comprising a dense central hub 
connected to at least four converging filaments (F1--F4) detected in absorption. 
Beyond this structure, numerous H$_2$ protostellar jets/outflows are also identified in the surrounding region. 

\item The HFS is kinematically confirmed by ATOMS H$^{13}$CO$^{+}$(1--0) emission over the velocity range [$-$19.2, $-$16.4] km s$^{-1}$.

\item An analysis of the H$^{13}$CO$^{+}$(1--0) integrated-intensity emission reveals molecular skeletons that closely trace the major JWST NIR-dark structures. The velocity-coded skeletons show steep velocity gradients indicative of inflow. This analysis supports the association of the NIR-dark features with molecular filaments.

\item The filaments (F1--F4) exhibit mass accretion rates of $7.6 \times 10^{-5}$ to $1.1 \times 10^{-4}$ M$_\odot$ yr$^{-1}$ and linewidths that increase from the outer regions toward the hub, with velocity gradients of $\sim$4--6 km s$^{-1}$ pc$^{-1}$. These trends are consistent with enhanced dynamical activity and gravity-driven turbulence toward the central region. 

\item No clear preferred alignment is found between the velocity gradients and the gravitational-force directions, which may indicate that the system is dynamically evolved and that multiple dynamical processes contribute to the observed gas motions.

\item G286c1, which hosts a protocluster and is associated with compact 1.3 GHz radio continuum emission and ALMAGAL cores, is concentrated toward the central hub of the HFS.

\item The radial distribution of ALMA 1.3 mm continuum cores (ALMAGAL and \citet{cheng20b}), measured from the hub center, reveals systematic radial trends. The core number density ($\rho$ [pc$^{-2}$]), surface density ($\Sigma$ [g cm$^{-2}$]), and core mass ($M_{\rm core}$ [M$_\odot$]) exhibit approximate power-law ($Y \propto r^{\alpha}$) trends, with slopes of $\alpha_\rho$ $\sim$$-2.8$ to $-1.1$, $\alpha_\Sigma$ $\sim$$-0.4$ to $-0.6$, and $\alpha_M$ $\sim$$-0.7$, while core diameters ($D_{\rm core}$ [AU]) remain nearly constant. The key result is the enhanced mass and core concentration toward G286c1, which is qualitatively consistent with a competitive-accretion scenario.

\item Strong 1.3 GHz free-free emission around G286c1 may contaminate the 1.3 mm continuum emission of nearby ALMAGAL cores, introducing systematic uncertainties in their derived masses and other continuum-based properties.

\item On the basis of the detection of PAH, H$_{2}$, and Br$\alpha$ using the \emph{JWST} facility, prominent 
PDRs are identified around the H\,{\sc ii}~region~A, which is powered by a B-type star. 
CHaMP $^{13}$CO position-velocity analysis indicates an expanding H\,{\sc ii} region.

\item The origin of the G286 HFS likely involves multiple coupled processes, including interactions between gas layers 
with different velocities and compression driven by feedback from the B-type star powering H\,{\sc ii}~region~A.
\end{itemize}

Overall, massive star and cluster formation in G286 appears to be regulated by the combined influence of filamentary gas transport, competitive core growth toward the hub, and stellar feedback. The molecular morphology and kinematics support organized gas motions within the HFS, while the complex velocity structure cautions against interpreting the observed gradients uniquely as gravitational inflow.

\section*{Acknowledgments}
We sincerely thank the anonymous referee for a careful review, whose constructive suggestions have led to significant improvements in this manuscript. The research work at Physical Research Laboratory is funded by the Department of Space, Government of India. NKB acknowledges the support of the China Postdoctoral Science Foundation through grant No. 2025M773187. RKY gratefully acknowledges the support from the Fundamental Fund of Thailand Science Research and Innovation (TSRI, Confirmation No. FFB680072/0269) through the National Astronomical Research Institute of Thailand (Public Organization). This work was supported by JSPS KAKENHI Grant Number JP26KF0052. This research was carried out in part at the Jet Propulsion Laboratory, California Institute of Technology, under contract with the National Aeronautics and Space Administration (80NM0018D0004). DA acknowledges Nazarbayev University FCDRG No.201223FD8821. CWL is supported by the Korea Astronomy and Space Science Institute grant funded by the Korea government (MSIT; Project No. 2025-1-841-02). This paper makes use of the following ALMA data: ADS/JAO.ALMA\#2019.1.00685.S. ALMA is a partnership of ESO (representing its member states), NSF (USA) and NINS (Japan), together with NRC (Canada), MOST and ASIAA (Taiwan), and KASI (Republic of Korea), in cooperation with the Republic of Chile. The Joint ALMA Observatory is operated by ESO, AUI/NRAO and NAOJ.
This research made use of {\it Astropy}\footnote[1]{http://www.astropy.org}, a community-developed core Python package for Astronomy \citep{astropy13,astropy18}. For figures, we have used {\it matplotlib} \citep{Hunter_2007} and IDL software.  
\subsection*{Data availability}
The {\it Spitzer} data and CHaMP $^{13}$CO line data underlying this article are available from the publicly accessible NASA/IPAC infrared science archive\footnote[2]{https://irsa.ipac.caltech.edu/frontpage/}.
The SARAO MeerKAT 1.3 GHz continuum data underlying this article are available from the publicly accessible website\footnote[3]{https://archive-gw-1.kat.ac.za/public/repository/10.48479/3wfd-e270/index.html}.
The ATOMS molecular line data and continuum data (ADS/JAO.ALMA\#2019.1.00685.S) underlying this article are available from the publicly accessible ALMA science archive\footnote[4]{https://almascience.nrao.edu/aq/}.
The ALMAGAL 1.3 mm continuum map underlying this article is available from the publicly accessible 
website\footnote[5]{https://www.almagal.org/index.php/en/almagal-browser}.
The ALMAGAL 1.3 mm continuum sources underlying this article are available from the publicly accessible 
website\footnote[6]{http://vizier.cds.unistra.fr/viz-bin/VizieR?-source=J/A+A/696/A151}.
The JWST images (DOI: https://doi.org/10.17909/0d6v-3y92) underlying this article are available
from the publicly accessible MAST archive \footnote[7]{https://archive.stsci.edu/missions-and-data/jwst}.
%


\bibliographystyle{mnras}
\bibliography{reference} 
%
%
\appendix
\section{Velocity channel map and H$_2$ column density estimations}
\label{sec:columndensity}
\begin{figure*}
\center 
\includegraphics[width=1\textwidth]{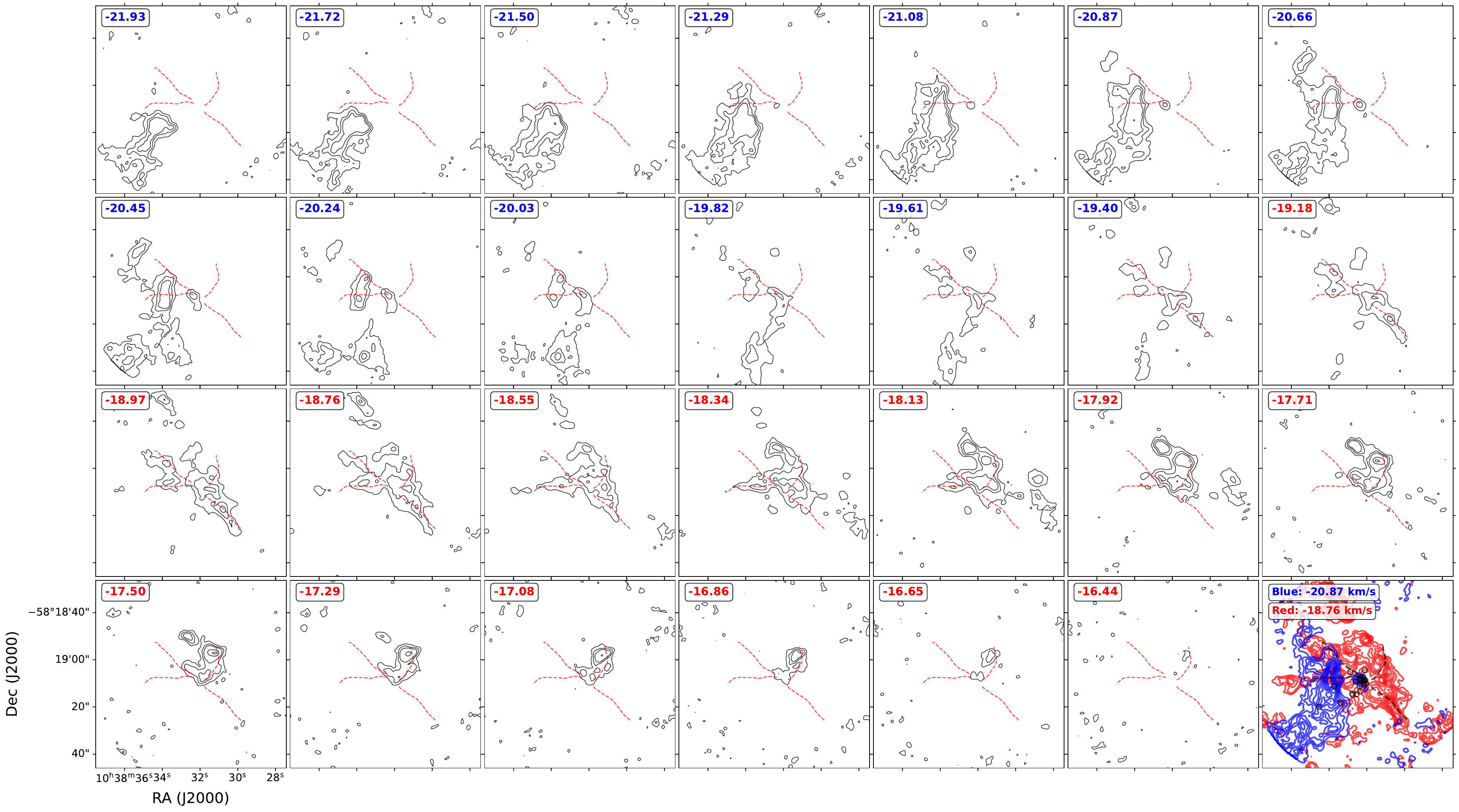}
\caption{Channel maps of H$^{13}$CO emission are shown with filament spine footprints overlaid in red (see also Figure~\ref{fig:hfs}). 
The velocity to each channel is indicated, with red labeling from $-$19.18 to $-$16.44 km s$^{-1}$ used to identify the hub-filament system morphology, and blue labeling from $-$21.93 to $-$19.40 km s$^{-1}$ for the blue-shifted components. 
Contours are plotted at [2, 4, 6, 12, 15, 18, 24] $\times$ the rms noise (in each channel), 
which varies from $9.98\times10^{-3}$ to $2.35\times10^{-2}$~Jy~beam$^{-1}$. 
In the bottom-right panel, ATOMS H$^{13}$CO$^{+}$ emission contours at $-$18.76 km s$^{-1}$ (red) and $-$20.87 km s$^{-1}$ (blue) are 
shown along with filament spine footprints. The red contours correspond to 0.149 $\times$ [0.1, 0.2, 0.3, 0.4, 0.5, 0.6, 0.7, 0.8, 0.9, 0.95] Jy beam$^{-1}$, 
while the blue contours are 0.223 $\times$ [0.1, 0.2, 0.3, 0.4, 0.5, 0.6, 0.7, 0.8, 0.9, 0.95] Jy beam$^{-1}$. 
Circles are the same as in Figure~\ref{fig4}a.}
\label{cfig:channelmap}
\end{figure*}

The ALMA H$^{13}$CO$^{+}$(1--0) line data were used to produce channel maps and a molecular hydrogen column density ($N(\mathrm{H_2})$) map.
Figure~\ref{cfig:channelmap} display the H$^{13}$CO$^{+}$(1--0) channel maps spanning velocities from $-$19.2 to $-$16.4 km s$^{-1}$. 
In order to highlight the HFS, each channel map is overlaid with filament spines that were visually identified from the \emph{JWST} images. 
The declination--velocity and right ascension--velocity of the ATOMS H$^{13}$CO$^{+}$ emission reveal at least two prominent velocity 
components at $-18.76$ and $-20.87$ km s$^{-1}$ (not shown), which appear connected in velocity space, suggesting a physical connection rather 
than a line-of-sight superposition. The bottom-right panel of Figure~\ref{cfig:channelmap} displays these components (red and blue contours), 
along with the distribution of ALMAGAL compact sources and filament spines. Most ALMAGAL sources lie within the overlapping region of the 
two velocity components, coinciding with the central hub of the HFS. 

The molecular hydrogen column density, $N(\mathrm{H_2})$, was estimated from the H$^{13}$CO$^{+}$ line emission integrated 
across the velocity interval of [$-$19.2, $-$16.4] km s$^{-1}$, assuming local thermodynamic equilibrium (LTE) and optically thin conditions. 
The H$^{13}$CO$^+$ column density can be computed using Equation 80 from \citealt{MangumShirley2015}:

\begin{equation}
N_{\rm tot} = \frac{8\pi k\nu^2}{hc^3 A_{ul}} \frac{Q(T_{\rm ex})}{g_u} \frac{\exp(E_u/kT_{\rm ex})}{\exp(h\nu/kT_{\rm ex})-1}\times \frac{W}{J_\nu(T_{\rm ex}) - J_\nu(T_{\rm bg})}
\label{eq:ntot_full}
\end{equation}

where $W$ is the integrated brightness temperature (K\,km\,s$^{-1}$), 
$\nu$ is the transition rest frequency, $A_{ul}$ is the Einstein coefficient, 
$g_u$ is the upper-level degeneracy, $E_u$ is the upper-level energy, 
$T_{\rm ex}$ is the excitation temperature, $Q(T_{\rm ex})$ is the rotational 
partition function, and $J_\nu(T) = (h \nu / k) / [\exp(h \nu / k T)-1]$ is the 
Rayleigh-Jeans equivalent temperature. The background corresponds to cosmic microwave background temperature, $T_{\rm bg}=2.73$~K.

For the H$^{13}$CO$^{+}$ $(1$--$0)$ transition, a rest frequency of $\nu = 86.7542884$~GHz, an Einstein $A$ coefficient of $A_{ul} = 3.85 \times 10^{-5}$~s$^{-1}$, an upper-level degeneracy of $g_u = 3$, and an upper-level energy of $E_u/k = 4.16$~K are adopted. The partition function is approximated as $Q(T_{\rm ex}) \simeq k T_{\rm ex} / (h B)$ under the condition $k T_{\rm ex} \gg h B$, where the rotational constant is $B = 43.377$~GHz \citep{CDMS2005}.

The observed moment--0 map, expressed in units of $\mathrm{Jy\,beam^{-1}\,km\,s^{-1}}$, is converted to brightness temperature using the standard Rayleigh--Jeans relation:

\begin{equation}
W(\mathrm{K\,km\,s^{-1}}) = 1.222 \times 10^{6} 
\frac{W(\mathrm{Jy\,beam^{-1}\,km\,s^{-1}})}{\nu^2(\mathrm{GHz}^2)\, 
\theta_{\rm maj}('')\, \theta_{\rm min}('') },
\end{equation}

where $\theta_{\rm maj}$ and $\theta_{\rm min}$ are the major and minor 
axes of the synthesized beam.

An excitation temperature of $T_{\rm ex} = 10$~K is adopted, consistent with values typically inferred for dense gas tracers in cold filamentary structures \citep[e.g.,][]{Pineda2010}. Using the parameters above, Equation~\ref{eq:ntot_full} is applied to each pixel of the moment-0 map to produce a H$^{13}$CO$^+$ column density map. The molecular hydrogen column density is then derived assuming a constant fractional abundance $X(\mathrm{H^{13}CO^+})  \sim1.7\times10^{-11}$ \citep[e.g.,][]{Maruta2010}. 
The resulting $N(\mathrm{H_2})$ map is shown in Figure~\ref{fig5}c. Assuming a cylindrical filament with a length of 0.25 pc and a width of 0.1 pc, the derived mass range of 19--39~$M_{\odot}$ implies a mean volume density of $n_{\rm H_2} \sim (1.4-2.9) \times 10^{5}$~cm$^{-3}$. This value is comparable to the critical density of the H$^{13}$CO$^{+}$ (1--0) transition ($\sim 10^{5}$~cm$^{-3}$), supporting the validity of the LTE approximation adopted in this work.
\section{Filamentary Structures in \emph{JWST} images}
\label{xxsec:jwstmap}
\begin{figure*}
	\center
	\includegraphics[width=\textwidth]{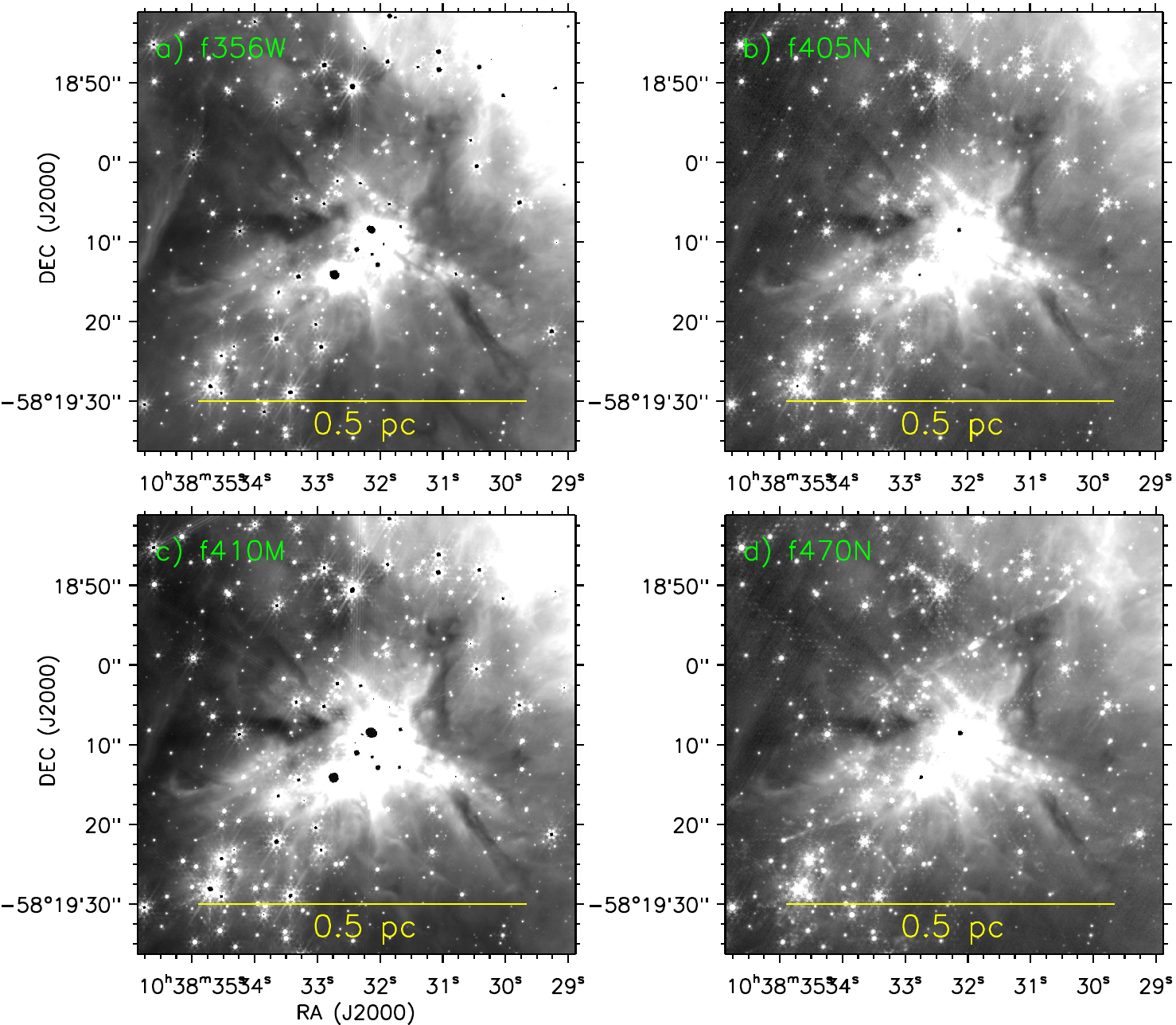}
	\caption{\emph{JWST} images of the same field shown in Figure~\ref{fig5}a obtained in the a) f356W, b) f410M, c) f405N, and d) f470N filters.}
	\label{fgx4}
\end{figure*}

Figure~\ref{fgx4} presents the \emph{JWST} f356W, f405N, f410M, and f470N images of the region shown in Figure~\ref{fig5}a (see also the dot-dashed box in Figure~\ref{fig4}a). Prominent filamentary absorption features are consistently detected in all four \emph{JWST} images, indicating that these embedded structures are robustly traced over the observed wavelength range. These filaments appear to converge toward the central hub-like region, revealing a network of dark filamentary structures seen in silhouette against the bright 
infrared nebulosity. The filamentary network is most prominent in the f356W and f410M images. The persistence of the same morphology across all filters confirms that the observed dark lanes represent dense dust-bearing filaments rather than filter-specific emission features. 
%
%
\section{CHaMP $^{13}$CO(1--0) line data}
\label{Asec:coem}
\begin{figure*}
\center
\includegraphics[width=\textwidth]{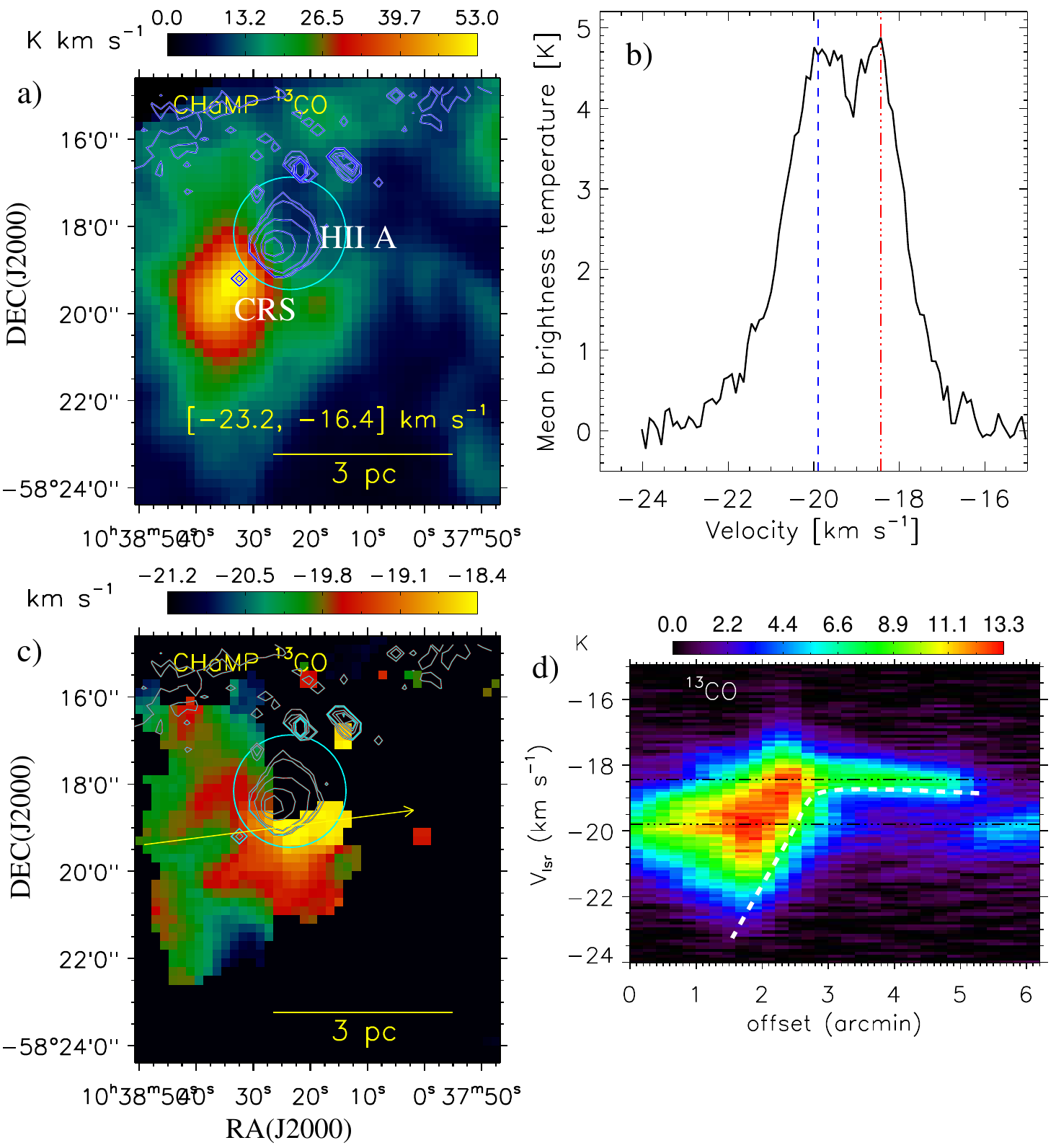}
\caption{a) SMGPS 1.3 GHz continuum contours (blue) overlaid on the 
CHaMP $^{13}$CO moment-0 map integrated over [$-$23.2, $-$16.4] km s$^{-1}$. 
b) Mean $^{13}$CO spectrum extracted from the circular region indicated in Figure~\ref{fig:apx1}a. 
c) SMGPS 1.3 GHz continuum contours (cyan) overlaid on the $^{13}$CO moment-1 map. 
d) Position-velocity diagram along the arrow shown in Figure~\ref{fig:apx1}c, with a 
dashed white curve tracing an arc-like structure.
In panels ``a'' and ``c'', contours are at 0.075, 0.15, 0.55, 2, 3, and 3.6 mJy beam$^{-1}$; 
the scale bar corresponds to 2.5 kpc. 
In panels ``b'' and ``d'', dot-dashed lines indicate velocities of $-$18.4 and $-$19.89 km s$^{-1}$.}
\label{fig:apx1}
\end{figure*}

This section analyzes CHaMP $^{13}$CO(1--0) line data toward G286 within [$-$23.2, $-$16.4] km s$^{-1}$. 
Figures~\ref{fig:apx1}a,~\ref{fig:apx1}b, and~\ref{fig:apx1}c display the integrated intensity map (or moment-0 map), 
mean $^{13}$CO($J$ = 1--0) profile (from the marked circular region in Figure~\ref{fig:apx1}a), and intensity weighted velocity map (or moment-1 map), respectively. 
The SMGPS 1.3 GHz continuum emission contours are also overlaid on both the moment maps, tracing ionized emission (i.e., CRS and H\,{\sc ii}~region A). 
Figure~\ref{fig:apx1}d presents the position-velocity diagram of $^{13}$CO($J$ = 1--0) along the path indicated by the arrow in Figure~\ref{fig:apx1}c. 
The dashed white curve tracing an arc-like structure is indicated in the position-velocity diagram (see Figure~\ref{fig:apx1}d). 
In Figures~\ref{fig:apx1}b and~\ref{fig:apx1}d, dot-dashed lines mark the velocities at $-$18.4 and $-$19.89 km s$^{-1}$. 
The $^{13}$CO($J$ = 1--0) line data reveal a noticeable velocity gradient toward the G286-clump, along with two velocity components separated by $\sim$1.5 km s$^{-1}$. 
Together with the arc-like structure in velocity space, these characteristics are consistent with a possible expanding shell, 
potentially associated with H\,{\sc ii}~region A.

\end{document}